\documentclass[pdflatex,sn-mathphys,Numbered]{sn-jnl}
\usepackage{graphicx}
\usepackage{multirow}
\usepackage{amsmath,amssymb,amsfonts}
\usepackage{amsthm}
\usepackage{mathrsfs}
\usepackage[title]{appendix}
\usepackage{xcolor}
\usepackage{textcomp}
\usepackage{manyfoot}
\usepackage{booktabs}
\usepackage{algorithm}
\usepackage{algorithmicx}
\usepackage{algpseudocode}
\usepackage{listings}
\numberwithin{equation}{section}
\usepackage{hyperref}

\newcommand{\beq}{\begin{equation}}
\newcommand{\eeq}{\end{equation}}
\newcommand{\beqa}{\begin{eqnarray}}
\newcommand{\eeqa}{\end{eqnarray}}

\newcommand{\be}{\begin{equation}}
\newcommand{\ee}{\end{equation}}
\newcommand{\bea}{\begin{eqnarray}}
\newcommand{\eea}{\end{eqnarray}}

\newcommand{\abs}[1]{\vert#1\vert}
\newcommand{\dd}{{\rm d}}
\newcommand{\e}{{\rm e}}
\newcommand{\etas}{\eta^{\star}}
\newcommand{\even}{{\rm even}}
\newcommand{\frad}[2]{\displaystyle{\displaystyle#1\over\displaystyle#2}}
\newcommand{\ii}{{\rm i}}
\renewcommand{\max}{{\rm max}}
\renewcommand{\min}{{\rm min}}
\newcommand{\mean}[1]{\langle#1\rangle}
\newcommand{\odd}{{\rm odd}}
\newcommand{\prob}{\mathbb{P}}
\newcommand{\reg}{{\rm reg}}
\newcommand{\sg}{{\rm sing}}
\newcommand{\til}[1]{{\widetilde{#1}}}

\newcommand{\xs}{x^{\star}}

\renewcommand{\AA}{{\mathcal{A}}}
\newcommand{\BB}{{\mathcal{B}}}
\newcommand{\CC}{{\mathcal{C}}}
\newcommand{\Int}{\mathop{{\rm Int}}}

\renewcommand{\Re}{\mathop{{\rm Re}}}

\newcommand{\eps}{{\varepsilon}}
\newcommand{\lam}{{\lambda}}
\renewcommand{\th}{{\theta}}
\newcommand{\lap}[1]{\mathop{\mathcal{L}}\limits_{#1}}

\newcommand{\Az}{{\mathbf{A}}}
\newcommand{\Bz}{{\mathbf{B}}}

\begin{document}

\title{On the first positive position of a random walker}

\author*[]{\fnm{Claude} \sur{Godr\`eche*}}\email{claude.godreche@ipht.fr}

\author[]{\fnm{Jean-Marc} \sur{Luck}}\email{jean-marc.luck@ipht.fr}

\affil[]
{\orgdiv{Universit\'e Paris-Saclay, CEA, CNRS},
\orgname{Institut de Physique Th\'eorique},
\postcode{91191}
\city{Gif-sur-Yvette},
\country{France}}

\abstract{
The distribution of the first positive position reached by a random walker starting from the origin
is fundamental for understanding the statistics of
extremes and records in one-dimensional random walks.
We present a comprehensive study of this distribution,
focusing particularly on its moments and asymptotic tail behaviour,
in the case where the step distribution is continuous and symmetric,
encompassing both diffusive random walks and L\'evy flights.}


\maketitle

\tableofcontents

\section{Introduction and summary}
\label{sec:intro}

This paper revisits the classic problem of characterising the distribution of the first positive sum
in a sequence of independent and identically distributed (iid) random variables, or
equivalently of the first positive position reached by a one-dimensional random walker starting
from the origin.
If the walker's steps are denoted by $\eta_1, \eta_2, \dots, \eta_n$,
its position $x_n$ at time $n$ is given by
\beq\label{eq:ma}
x_n=x_{n-1}+\eta_n,\qquad x_0=0.
\eeq
Let $N\ge1$ denote the first time when the walker's position is positive,
and $H=x_N$ denote this first positive position (see figure~\ref{fig:sketch}).
How can we characterise the joint distribution of the two random variables $(N,H)$?

This problem, along with related questions,
has been the focus of significant mathematical research since the 1950s,
led by prominent figures such as Spitzer~\cite{spitzer,spitzer1,spitzer2,spitzer3},
Pollaczek~\cite{pollaczek}, Feller~\cite{feller2}, Blackwell~\cite{blackwell} and
Baxter~\cite{baxter}.
By the end of the 1950s, the essential aspects of the problem were understood.
The techniques employed were primarily probabilistic and combinatorial,
though influenced by earlier analytical developments known as the Wiener-Hopf approach~\cite{hopf}%
\footnote{See~\cite{noble} for an overview of the Wiener-Hopf approach and~\cite{LA} for a
historical account.}.
Notably, two of Spitzer's papers on this topic are titled \textit{The Wiener-Hopf equation whose
kernel is a probability density}~\cite{spitzer1,spitzer2}.
The primary motivation of these two papers was to investigate the maximum position reached by a
random walk after $n$ steps.
In the second of these papers, Spitzer provides a summary of his results, which we outline below
in order to set the context for our study.
The key equation of the whole approach is the homogeneous Wiener-Hopf integral equation,
\beq\label{eq:GSpitz}
G(x)=\int_{0}^{\infty}\dd y\, G(y)\rho(x-y)\qquad(x>0),
\eeq
with boundary condition $G(0)=1$,
where the kernel $\rho(x)$ is the probability density of the walker's steps,
\beq
\rho(x)=\frac{\dd}{\dd x}\prob(\eta<x).
\eeq
Apart from the assumption that the step distribution is symmetric (i.e., even) and continuous,
no other restrictions apply to it.
Thus, the process defined in~(\ref{eq:ma}) describes diffusive random walks
if the diffusion constant $D$, defined as
\be
2D=\mean{\eta^2}=\int_{-\infty}^{\infty}\dd x\,x^2 \rho(x),
\label{ddef}
\ee
is finite, and L\'evy flights if it is infinite.
The integral equation~(\ref{eq:GSpitz}) has a unique solution, whose derivative,
$g(x)=G'(x)$,
is given in Laplace space, for $\Re p>0$, by
\beq
\label{eq:Poll-S1}
\hat g(p)
=\int_0^\infty\dd x\, \e^{-p x} g(x)
=\exp\left(-\frac{p}{\pi}\int_0^\infty\frac{\dd q }{p^2+q^2}\ln(1-\til \rho(q))\right),
\eeq
where
\beq\label{eq:four}
\til \rho(q)=\int_{-\infty}^{\infty}\dd x\, \e^{\ii q x} \rho(x)
\eeq
is the Fourier transform of the density $\rho(x)$.
The integral representation~(\ref{eq:Poll-S1}) for $\hat g(p)$ in terms of $\til \rho(q)$ is known
as the Pollaczek-Spitzer formula.

For large $x$, $G(x)$ is such that
\beq\label{eq:meanHD}
\lim_{x\to\infty}\frac{G(x)}{x}=\frac{1}{\sqrt{D}},
\eeq
where $D$ may be finite or infinite.
In the former case, writing
\beq\label{eq:deflam}
G(x)=\frac{x+\lam(x)}{\sqrt{D}},
\eeq
the function $\lam(x)$ has a finite limit as $x\to\infty$, given by
\beq\label{eq:l-Milne}
\ell=-\frac{1}{\pi}\int_0^\infty\frac{\dd q}{q^2}\ln\frac{1-\til\rho(q)}{D q^2},
\eeq
whenever $\mean{\abs{\eta}^3}$ is finite.
Thus
\beq\label{eq:homo1}
G(x)\approx\frac{x+\ell}{\sqrt{D}},
\eeq
where $\ell$ is known as the extrapolation length.

Spitzer also provides a probabilistic interpretation of these results, which we will explore in
detail.
This interpretation relates the solution $G(x)$ of the homogeneous equation~(\ref{eq:GSpitz})
to our object of study, the distribution of the first positive sum $H$ (or the first positive
position of the walker):
\beq\label{eq:fH}
f(x)=\frac{\dd}{\dd x}\prob(H<x).
\eeq
In particular, he derives the relation
\beq\label{eq:baxter}
\hat g(p)=\frac{1}{1-\hat f(p)}\qquad(\Re p>0),
\eeq
where
\be
\hat f(p)=\mean{\e^{-p H}}=\int_0^{\infty}\dd x\, \e^{-p x} f(x).
\ee
As a consequence of~(\ref{eq:baxter}), the analysis of~(\ref{eq:Poll-S1}) to leading order at small
$p$ yields the mean first positive position
\beq
\mean{H}=\sqrt{D}.
\label{h1res}
\eeq
All the results recalled above are central to the study we shall undertake.
In particular, the identity~(\ref{eq:baxter}), attributed to Baxter by Spitzer,
plays a fundamental role in what follows.

As mentioned earlier, while most advances on the probabilistic side were achieved during the 1950s,
articles on the subject have continued to appear regularly in the mathematical literature since then
(see, e.g.,~\cite{sinai,rogo1,rogo2,rogo7,lai,doney,gru,peres,alev,nagaev}).
More recently, the distribution of the pair $(N,H)$,
and especially that of the first positive position $H$,
has sparked renewed interest within the statistical physics community.
This interest stems from the central role these random variables play
in studying the statistics of extremes and records in random walks and L\'evy flights
(see~\cite{gmsprl}, and~\cite{revue,msbook} for reviews).
The theory of records for random walks represents a natural step in complexity beyond the classical
theory of records,
which is based on sequences of iid random
variables~\cite{chandler,renyi,glick,arnold,nevzorov,nevzorov2,bunge}.
Records for sequences of iid random variables are naturally encountered in statistical mechanics
models (see, e.g.,~\cite{redner,krug,glrecord,doussal,miller,ben}).
More broadly, the study of records has attracted significant attention due to its wide-ranging
applications in complex systems (see~\cite{revue,msbook} for references).

This work aims to shed new light, from various perspectives,
on the distribution of the first positive position of a one-dimensional random walker,
presenting an essentially self-contained exposition.

In section~\ref{sec:outline} we revisit the main outlines of the probabilistic formalism underlying
the results sketched above.
Our primary purpose is to clarify the probabilistic significance of quantities
such as $G(x)$, $g(x)$ and $f(x)$, and of the relationships between them.
This interpretation of the results mentioned above is enriched with concepts from the
Wiener-Hopf method, particularly drawing on developments by Feller~\cite{feller2}.
According to Feller, although the mathematical apparatus of the Wiener-Hopf method is not
essential, its underlying concepts are, and they lend themselves to a probabilistic interpretation.

The following sections present new results.
We begin, in section~\ref{sec:asympt}, with a systematic investigation of the asymptotic tail behaviour
of the distribution of $H$ near its upper edge, whether finite or infinite.
The analysis, which is based on the Wiener-Hopf factorisation identity~(\ref{eq:fellerfactor}),
addresses three classes of step distributions.

The first class encompasses all step distributions $\rho(x)$ whose Laplace transform~$\hat\rho(p)$
is analytic over the entire complex $p$-plane.
This includes, on the one hand, distributions with finite support (e.g., the uniform distribution)
and, on the other hand, step distributions that extend to infinity and decay faster than any
exponential function (e.g., the Gaussian distribution).
We show that the distributions of $\eta$ and $H$ are asymptotically equivalent:
$f(x)\approx \rho(x)$,
as $x$ approaches the upper edge of the support of $f(x)$, whether finite or infinite.

The second class includes step distributions whose decay is essentially exponential and for which
the Laplace transform $\hat\rho(p)$ is analytic within a strip.
There, the tails of both distributions satisfy $f(x)\approx K \rho(x)$,
where the proportionality constant $K$ depends on details of $\rho(x)$.

The third class consists of step distributions~$\rho(x)$ with subexponential decay, i.e., those
whose falloff is slower than any exponential function,
so that only the Fourier transform $\til \rho(q)$ is well-defined.
In the case where $\rho(x) \approx c \abs{x}^{-(1+\theta)}$,
we find $f(x)\approx ax^{-(1+\sigma)}$,
where the tail exponent $\sigma$ equals $\th-1$ for diffusive walks ($\theta>2$), and $\th/2$ for
L\'evy flights ($0<\theta<2$).
Figure~\ref{alphaplot} illustrates the dependence of the tail exponent $\sigma$ on $\theta$.
The regime where $\sigma=\th/2$ was identified by Sinai~\cite{sinai} and revisited in~\cite{gmsprl},
whereas the regime where $\sigma=\th-1$ is original to this work.
The amplitude $a$ has distinct expressions in these two regimes, as given in~\eqref{agt}
and~\eqref{alt}.
We also consider step distributions falling off faster than any power law,
such as those with stretched exponential tails.
The asymptotic relationship between $f(x)$ and $\rho(x)$ is given
in~(\ref{resother}),~(\ref{othergal}).

Section~\ref{sec:moments} presents a detailed study of the moments of $H$.
Several attempts have been made in the past to resolve this question (see,
e.g.,~\cite{lai,doney,alev,nagaev}).
However, none are entirely satisfactory, as they fail to provide simple and systematic expressions
for the moments.
The methods used in this work greatly simplify the matter and yield more elegant expressions.
We assume that the step distribution $\rho(x)$ decreases rapidly, ensuring that all moments of $H$
are finite.
The moments of $H$ are expressed in terms of the cumulants of the excess length $E$, which is the
stationary limit of the overshoot $E_x$ of the random walk over the `barrier' located at $x$ (see
figure~\ref{fig:sketchx}).
The even cumulants of $E$ have explicit expressions in terms of the moments of the step distribution (see~\eqref{ceven}),
whereas the odd ones have more intricate integral representations (see~\eqref{codd}).
The first cumulant, $c_1=\mean{E}$, is identified with $\ell$, the extrapolation length.
Therefore, in some sense, the higher-order odd cumulants $c_3,c_5,\dots$ are generalisations of the latter.

Section~\ref{stable} is devoted to the class of stable step distributions,
with an index in the range $0<\alpha\le2$.
We successively consider the Gaussian distribution ($\alpha=2$),
the Cauchy distribution ($\alpha=1$) and general L\'evy stable distributions.
For the Gaussian case, we obtain explicit expressions of the moments of $H$ to arbitrary order,
in terms of the Riemann zeta function (see~\eqref{eq:momentgauss}).
We also derive a power-series expansion in~$x$ for the distribution $f(x)$, yielding an
accurate convergent series representation of the latter (see figure~\ref{gaussplot}).
For the Cauchy flight, we derive a closed-form expression for the distribution $f(x)$ in Laplace
space (see~\eqref{fcau}).
This allows the determination of the asymptotic behaviour of the distribution $f(x)$ at large $x$,
as well as an accurate calculation of the distribution $f(x)$
(see figure~\ref{cauchyplot}).
For L\'evy flights, corresponding to an arbitrary index $0 < \alpha < 2$, the asymptotic form of
the distribution $f(x)$ for large $x$, as well as its power-series expansion in $x$, can be derived
following the same approach as in the Gaussian case.

Section~\ref{sec:anal} presents a complementary analytical approach to the problem,
building on the longstanding observation
that Wiener-Hopf equations can be solved using elementary methods
when the Laplace transform~$\hat\rho(p)$ of the step density is a rational function of
$p$~\cite{wick,chandra}.
This approach bypasses the formal mathematical framework of the Wiener-Hopf method,
while explicitly reproducing its outcomes.
We are thus able to establish a range of general results, including Wiener-Hopf factorisation
properties, general expressions for $G(x)$, $g(x)$, and $f(x)$ in Laplace space, as well as an
independent derivation of~\eqref{eq:baxter}.
We then proceed with a detailed analysis of three representative examples:
the symmetric exponential distribution,
the double symmetric exponential distribution, and symmetric Erlang distributions.

The details of some derivations are deferred to an appendix.

\section{The main outlines of the probabilistic formalism}
\label{sec:outline}

In this section, we introduce the foundational concepts necessary for our study.
The objective is, starting from the definitions of a few basic events,
to give a probabilistic meaning to the quantities introduced above, such as $G(x)$ and $g(x)$,
and to explain the origin of~(\ref{eq:meanHD}) and~(\ref{eq:deflam}).
In section~\ref{sec:recap}, we will summarise the progress made and highlight what remains to be
addressed.
This will be supplemented by the introduction of the Wiener-Hopf factorisation identity and
completed from another perspective in section~\ref{sec:anal}.
Most of the material in this exposition is inspired by the works of Feller~\cite{feller2} and
Spitzer~\cite{spitzer1,spitzer2}.

\subsection{Ladder variables, records and the first positive position}

Let us come back to the definition of the random walk given in~(\ref{eq:ma}).
The steps $\eta_1, \eta_2, \dots$ are iid random variables with common probability density
$\rho(x)$.
The induced random walk is
\be
x_n=\eta_1 + \cdots + \eta_n, \qquad x_0=0.
\ee
As noted above, $\rho(x)$ being symmetric and continuous, but otherwise arbitrary,
this random walk may be either diffusive or a L\'evy flight.

\begin{figure}[!htbp]
\begin{center}
\includegraphics[angle=0,width=.6\linewidth,clip=true]{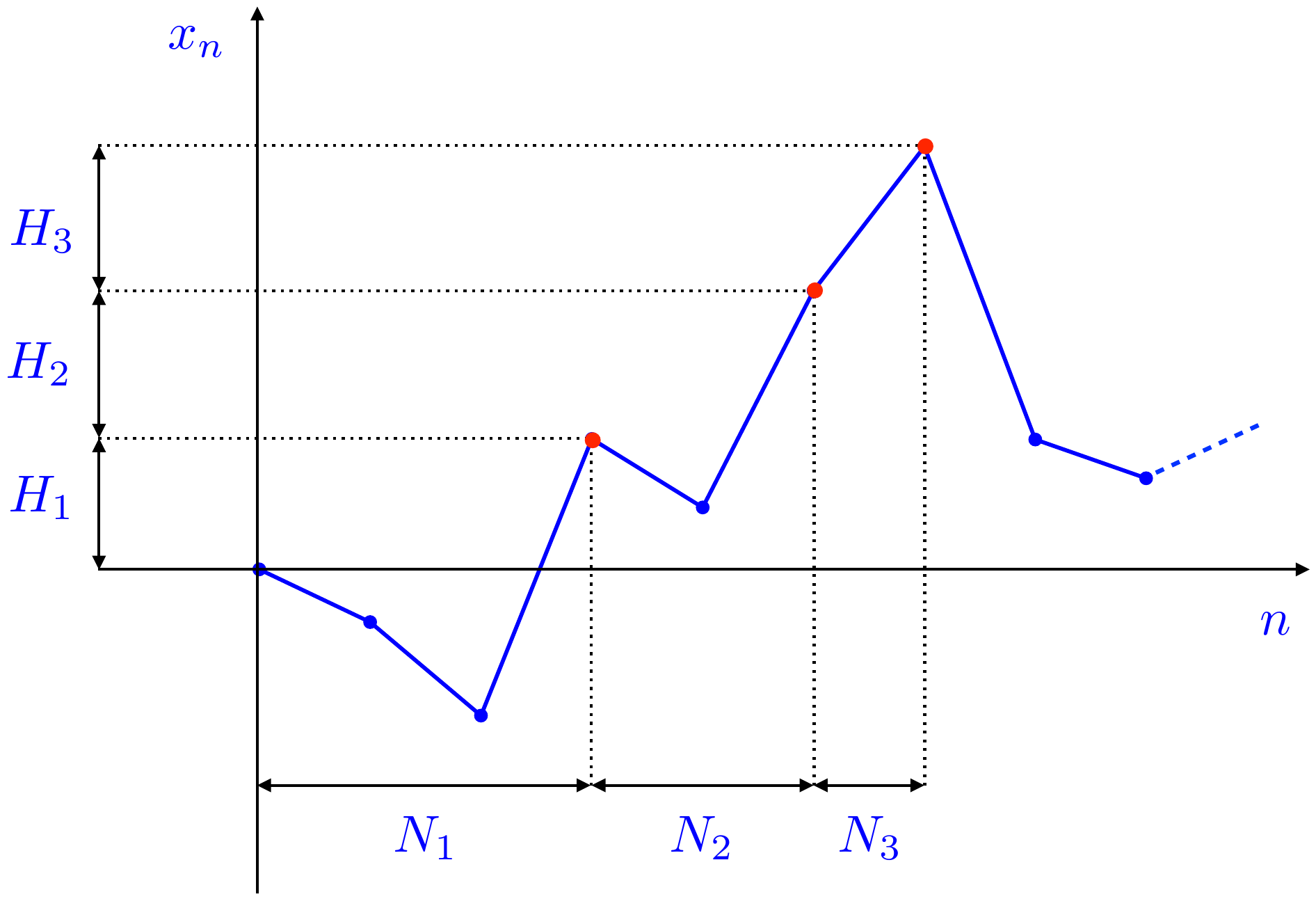}
\caption{The red dots represent the ladder points, or records, of the random walk.
The coordinates of the first ladder point, or first record, are denoted by $(N_1, H_1)$.
The successive ladder times, or record times, are denoted by $N_1$, $N_2$, $N_3$.
The successive ladder heights, or record heights, are denoted by $H_1$, $H_2$, $H_3$.
The ladder points form a two-dimensional renewal process, meaning that the walk starts afresh from
each red dot, taken as the new origin of the walk.}
\label{fig:sketch}
\end{center}
\end{figure}

As emphasised by Feller in~\cite[chapter XII]{feller2}, which is devoted to the one-dimensional
random walk,
{\it Looking at the graph of a random walk, one notices as a striking feature the points where $x_n$
reaches a record value, that is, where $x_n$ exceeds all previously attained values $x_0, \dots,
x_{n-1}$.
These are the ladder points [...].
The theoretical importance of ladder points derives from the fact that the sections between them
are probabilistic replicas of each other, and therefore important conclusions concerning the random
walk can be derived from a study of the first ladder point.}
These observations are illustrated in figure~\ref{fig:sketch}.

Let $(N,H)$ represent the coordinates of the first ladder point, or first record.
The first ladder epoch, or record time (that is, the time of the first entry into the positive
half-axis), is defined by the event\footnote{Since $\rho(x)$ is continuous, no distinction is made
between strict and non-strict inequalities.}
\be
\{N=n\}=\{x_1 < 0,\dots,\ x_{n-1} < 0,\ x_n > 0\}.
\ee
The first ladder height, or first record value (that is, the first positive position of the walker),
is given by $H=x_{N}$.
These quantities were originally introduced by Blackwell~\cite{blackwell}.

\subsection{Three fundamental events}

We introduce the following three events, defined for $n\ge1$ and $x>0$,
\beqa\label{eq:ABC}
\AA_n(x) &=& \{x_1 < 0,\dots,\ x_{n-1} < 0,\ 0 < x_n < x\},
\nonumber\\
\BB_n(x) &=& \{x_1 > 0,\dots,\ x_{n-1} > 0,\ 0 < x_n < x\},
\nonumber\\
\CC_n(x) &=& \{x_n > x_0,\ x_n > x_1,\dots, x_n > x_{n-1},\ 0 < x_n < x\}.
\eeqa
In essence, these events can be interpreted as follows:

\begin{enumerate}

\item[1.]
$\AA_n(x)$ describes a first-passage event, specifically the first entry into the interval $(0, x)$
at epoch $n$.

\item[2.]
$\BB_n(x)$ describes a survival event, specifically when the interval $(0, x)$ is visited
at epoch $n$ without prior passage through negative values.

\item[3.]
$\CC_n(x)$ describes the occurrence of a record at time $n$ within the interval $(0, x)$.

\end{enumerate}

All quantities considered below are based on these three events.
The corresponding probabilities are analysed successively in what follows.

\subsection{Event $\AA_n$(x) and definition of $F(x)$}

The probability of event $\AA_n(x)$, that is the probability of first entry in $(0,x)$ at epoch
$n$, represents the joint distribution of $N$ and $H$,
\beq\label{eq:Fn}
F_n(x)=\prob(\AA_n(x))=\prob(N=n,\ H<x)\qquad(n\ge 1,\ x>0).
\eeq
The marginal distribution of $H$ is deduced by summing over $n$,
\beq\label{eq:marginal}
F(x)=\prob(H < x)=\sum_{n \ge 1} F_n(x).
\eeq
The density associated to $F_n(x)$ is given by
\beq\label{eq:fndef}
f_n(x)=\frac{\dd}{\dd x}F_n(x),
\eeq
i.e.,\footnote{This density is referred to as $J(n,x)$ in~\cite{gmsprl}.
It also appears in the study of the order statistics of random walks~\cite{mounaix1,mounaix2}.}
\beq\label{eq:fndef2}
f_n(x)\dd x=\prob(x_1 < 0,\dots,\ x_{n-1} < 0,\ x < x_n < x+\dd x),
\eeq
where $x>0$.
This represents the probability that the walker, starting at the origin, stays below the origin up
to epoch $n -1$ and then makes a jump to the positive side, reaching $x > 0$ (up to $\dd x$) at
epoch $n$.
The density associated to $F(x)$,
\beq\label{eq:fdef}
f(x)=\sum_{n \ge 1} f_n(x)=\frac{\dd}{\dd x} F(x),
\eeq
is the central object of our study, that is the distribution of the first positive sum $H$ (or the
first positive position of the walker) (see~\eqref{eq:fH}).
The marginal distribution of $N$, or first-passage probability, is obtained by summation on $x$:
\beq\label{eq:deffn}
f_n=\prob(N=n)=\int_0^\infty \dd x\, f_n(x)
=\lim_{x\to\infty}F_n(x).
\eeq
For symmetric continuous step distributions, the first-passage probability $f_n$ and the survival
probability $g_n$ defined below (see~\eqref{eq:gndef}) have simple universal expressions, given by
Sparre Andersen theory, as will be recalled in section~\ref{sec:sparre}.

\subsection{Event $\BB_n(x)$ and definition of $G(x)$}
\label{sec:defG}

Following the same approach as above, we define the following quantities.
The probability of event $\BB_n(x)$ is denoted by
\beq\label{eq:defGn}
G_n(x)=\prob(\BB_n(x))\qquad(n \ge 1,\ x>0).
\eeq
This represents the probability that the random walk remains positive up to time $n-1$ and reaches
a position between $0$ and $x$ at time $n$.
For $n=0$, we set
\be
G_0(x)=\mathbb{I}(x\ge0),
\ee
where $\mathbb{I}(x\ge0)$ is equal to 1 if $x\ge0$ and to 0 otherwise.
The sum
\beq\label{eq:defG}
G(x)=\sum_{n\ge0}G_n(x)=\mathbb{I}(x\ge0)+\sum_{n\ge1}G_n(x)
\eeq
represents the expected number of visits to the interval $(0, x)$ before entering the region
$(-\infty, 0)$ (this is further discussed in section~\ref{sec:duality}).
The equivalence between the quantity defined in~(\ref{eq:defG}) and the solution of the homogeneous
equation~(\ref{eq:GSpitz}) will also be addressed later (see sections~\ref{sec:recap}
and~\ref{sec:GChandra}).

By differentiation of~(\ref{eq:defGn}) and~(\ref{eq:defG}), we obtain respectively the density
\be
g_n(x)=\frac{\dd}{\dd x} G_n(x),
\ee
such that
\beq\label{eq:gnxdef}
g_n(x) \dd x=\prob(x_1>0,\dots,\ x_{n-1}>0,\ x<x_n<x+\dd x),
\eeq
and the density
\beq\label{eq:gxdef}
g(x)=\frac{\dd}{\dd x} G(x)=\sum_{n \ge 0} g_n(x)=\delta(x)+g_\reg(x),
\eeq
with regular part
\beq\label{gregdef}
g_\reg(x)=\sum_{n\ge1}g_n(x).
\eeq
The density $g_n(x)$ satisfies the following recurrence,
which is a forward equation derived by conditioning on the last step,
\beq\label{eq:recurr}
g_n(x)=\int_0^\infty \dd y\, g_{n-1}(y) \rho(y - x)\qquad(n\ge1),
\eeq
with $g_0(x)=\delta(x)$.
As a consequence,
\beq\label{eq:integral}
g(x)=\delta(x) + \int_0^\infty \dd y \, g(y) \rho(x - y).
\eeq
By the very definitions of $f_n(x)$ and $g_n(x)$, we have
\be
f_n(0)=g_n(0) \qquad(n \ge 1).
\label{fgnzero}
\ee
The quantity
\beq
\omega=f(0)=g_\reg(0)=G'(0)
\label{omegadef}
\eeq
can be evaluated as the limit of the product $p\hat f(p)$ as $p\to+\infty$.
Using~(\ref{eq:Poll-S1}) and~(\ref{eq:baxter})
(or~(\ref{psone}) and~(\ref{ione})), we obtain the integral representation
\beq
\omega=-\frac{1}{\pi}\int_0^\infty\dd q\,\ln(1-\til\rho(q)),
\label{omegaint}
\eeq
which is similar to the expression~(\ref{eq:l-Milne}) for the extrapolation length $\ell$.
Since $G(x)$ is a probability, the product $\omega\sqrt{D}$ is dimensionless.
Moreover, in view of~(\ref{eq:homo1}), the same applies to the ratio $\ell/\sqrt{D}$.
We denote these two dimensionless quantities by
\beq
\Az=\frac{\ell}{\sqrt{D}},\qquad
\Bz=\omega\sqrt{D}=\frac{g_\reg(0)}{g_\reg(\infty)}.
\label{abdef}
\eeq

An alternative expression for $\omega$ can be obtained by expanding the logarithm
in~(\ref{omegaint}) in powers of $\til\rho(q)$, which yields
\beq
\omega=\sum_{n\ge1} \frac{f_{x_n}(0)}{n},
\label{omegareturns}
\eeq
where
\beq
f_{x_n}(x)=\frac{1}{2\pi} \int_{-\infty}^\infty \dd q\, \til\rho(q)^n\, \e^{-\ii q x}
\eeq
is the distribution of the walker's position $x_n$ at time $n$.
The identity
\be
g_n(0)=\frac{f_{x_n}(0)}{n}
\ee
actually holds for any discrete time $n$ and any continuous symmetric step distribution.

Integrating~\eqref{eq:gnxdef} on $x$ yields the survival probability of the walk\footnote{By
duality (see section~\ref{sec:duality}), $g_n$ is also the probability that $n$ is a ladder epoch.},
\beq
g_n\label{eq:gndef}
= \prob(x_1 > 0,\dots,\ x_n > 0)
= \int_0^\infty \dd x\, g_n(x)
=\lim_{x\to\infty}G_n(x),
\eeq
which is related to $f_n$ by $f_n=g_{n-1} - g_n$ for $n \ge 1$,
and so
\be
\sum_{n \ge 1} f_n=g_0=1,
\ee
ensuring the normalisation of the probabilities~(\ref{eq:deffn}).

Finally, the following relation holds~\cite{gmsprl,revue,msbook}, as is evident upon inspection:
\be
f_n(x)=\int_0^\infty\dd y\, g_{n-1}(y)\rho(y+x)
\qquad(n\ge1),
\ee
hence
\beq\label{eq:fondam1}
f(x)
=\int_0^\infty\dd y\, g(y)\rho(y+x),
\eeq
from which we deduce (see~(\ref{eq:gxdef}) and~(\ref{eq:integral}))
\beq\label{eq:newf0}
f(0)=\int_0^\infty\dd y\, g(y)\rho(y)=g_\reg(0),
\eeq
thus recovering~(\ref{omegadef}).

\subsection{Event $\CC_n(x)$ and renewal function $\Psi(x)$}

\begin{figure}[!htbp]
\begin{center}
\includegraphics[angle=0,width=.6\linewidth,clip=true]{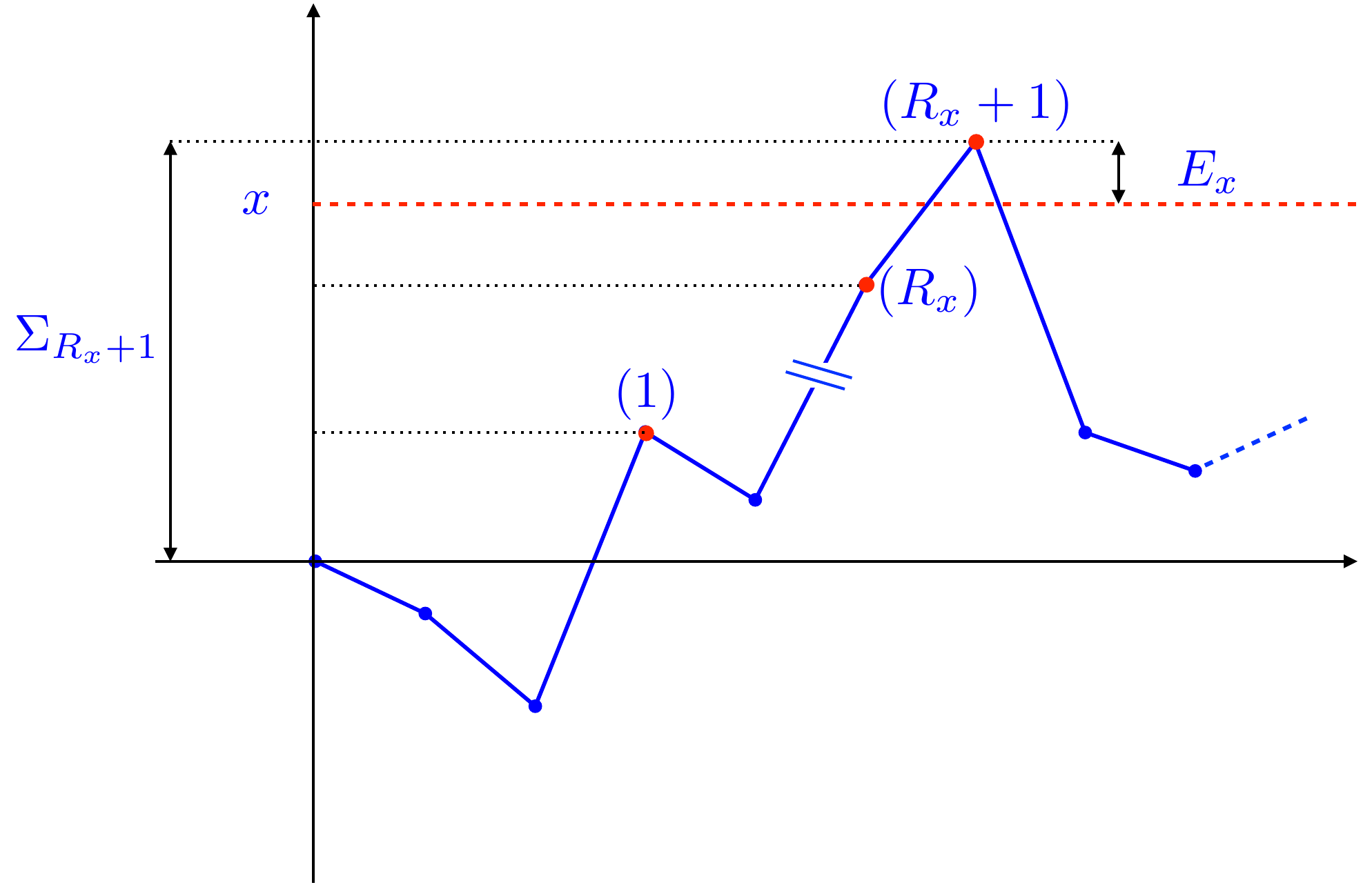}
\caption{The number of records of the random walk in the interval $(0,x)$ (not taking into account
the record at the origin) is denoted by $R_x$.
The excess length $E_x$ represents the overshoot of the walk over the `barrier' at $x$.
The corresponding height of the walk is denoted by $\Sigma_{R_x+1}$.
For a diffusive random walk, the mean of this quantity is equal to $\sqrt{D}\,G(x)$,
see~(\ref{eq:smith}) and~(\ref{eq:overshoot}).
At large $x$, $\mean{E_x}\to\ell$, the extrapolation length.}
\label{fig:sketchx}
\end{center}
\end{figure}

Consider the renewal function for the height process defined as~\cite{feller2}
\beq\label{eq:psiH}
\Psi(x)=\Psi_0(x)+\sum_{m\ge1}\prob(\Sigma_m<x),
\eeq
where
\beq
\Sigma_m=H_1+\cdots+H_m
\label{sigmadef}
\eeq
and
\be
\Psi_0(x)=\mathbb{I}(x\ge0).
\ee
The renewal function $\Psi(x)$ has three distinct interpretations:

\begin{enumerate}

\item[1.]
As a consequence of its very definition, it is given by the sum
\beq\label{eq:psi-2}
\Psi(x)=\Psi_0(x)+\sum_{n\ge1}\Psi_n(x),
\eeq
where
\be
\Psi_n(x)=\prob(\CC_n(x))\qquad(n\ge 1,\ x>0).
\ee

\item[2.]
Using the fact that
\be
\frac{\dd}{\dd x}\prob(\Sigma_m<x)=(f\ast)^m(x),
\ee
is an $m$-fold convolution,
it follows by differentiating~(\ref{eq:psiH}) that
\be
\psi(x)=\frac{\dd}{\dd x}\Psi(x)=\delta(x)+\sum_{m\ge1}(f\ast)^m(x).
\ee
This function is the renewal density of the height process,
such that $\psi(x)\dd x$ is the mean number of events (here, records) in the interval $(x,x+\dd x)$
(see~(\ref{eq:Nx})).
In Laplace space, we obtain
\beq\label{eq:id1}
\hat \psi(p)
=\frac{1}{1-\hat f(p)},
\eeq
where we denote the usual Laplace transforms as
\beqa
\hat \psi(p)&=&\lap{x}\psi(x)=\int_{0}^\infty\dd x\,\e^{-p x} \psi(x),
\nonumber\\
\hat f(p)&=&\lap{x}f(x)=\int_{0}^\infty\dd x\,\e^{-p x} f(x).
\eeqa
We shall also later require the bilateral Laplace transform of the density $\rho(x)$, denoted by
\beq\label{eq:bil}
\hat\rho(p)=\mean{\e^{-p\eta}}=\int_{-\infty}^\infty\dd x\,\e^{-p x} \rho(x).
\eeq

\item[3.]
Finally, we have
\beq\label{eq:psix}
\Psi(x)=1+\mean{R_x},
\eeq
where $R_x$ is the number of records in the interval $(0,x)$,
not taking into account the record at the origin (see figure~\ref{fig:sketchx}).
This can be understood as follows.
We can first check~(\ref{eq:psix}) by taking the Laplace transform of its differentiated form
and using the expression for $\lap{x}\mean{R_x}$ known from renewal theory
(see~(\ref{eq:meanN})).
This yields
\beq\label{eq:Nx}
\hat \psi(p)=p\,\lap{x}(1+\mean{R_x})=1+p\lap{x}\mean{R_x}=\frac{1}{1-\hat f(p)},
\eeq
which is precisely~(\ref{eq:id1}).
We can also observe that the following identity holds:
\beq
\mean{R_x}=\sum_{m\ge1}\prob(R_x \ge m)=\sum_{m\ge1}\prob(\Sigma_m<x),
\eeq
implying that the right-hand sides of~(\ref{eq:psiH}) and~(\ref{eq:psix}) coincide.

\end{enumerate}

\subsection{Duality and the equality of $G(x)$ and $\Psi(x)$}
\label{sec:duality}

Introduce the dual steps $\etas_1=\eta_n$, $\etas_2=\eta_{n-1},$ $\dots$, $\etas_n=\eta_1$.
Their partial sums are
\be
\xs_0=0,
\quad \xs_1=x_n-x_{n-1},
\quad\xs_k=x_n-x_{n-k},\quad \xs_n=x_n.
\ee
The dual walk, as defined, is obtained from the original by fixing the endpoint at the origin,
and then performing a $180^{\circ}$ rotation.
The joint distributions of $(x_1,\dots,x_n)$ and its dual are identical.

\begin{figure}[!htbp]
\begin{center}
\includegraphics[angle=0,width=.9\linewidth,clip=true]{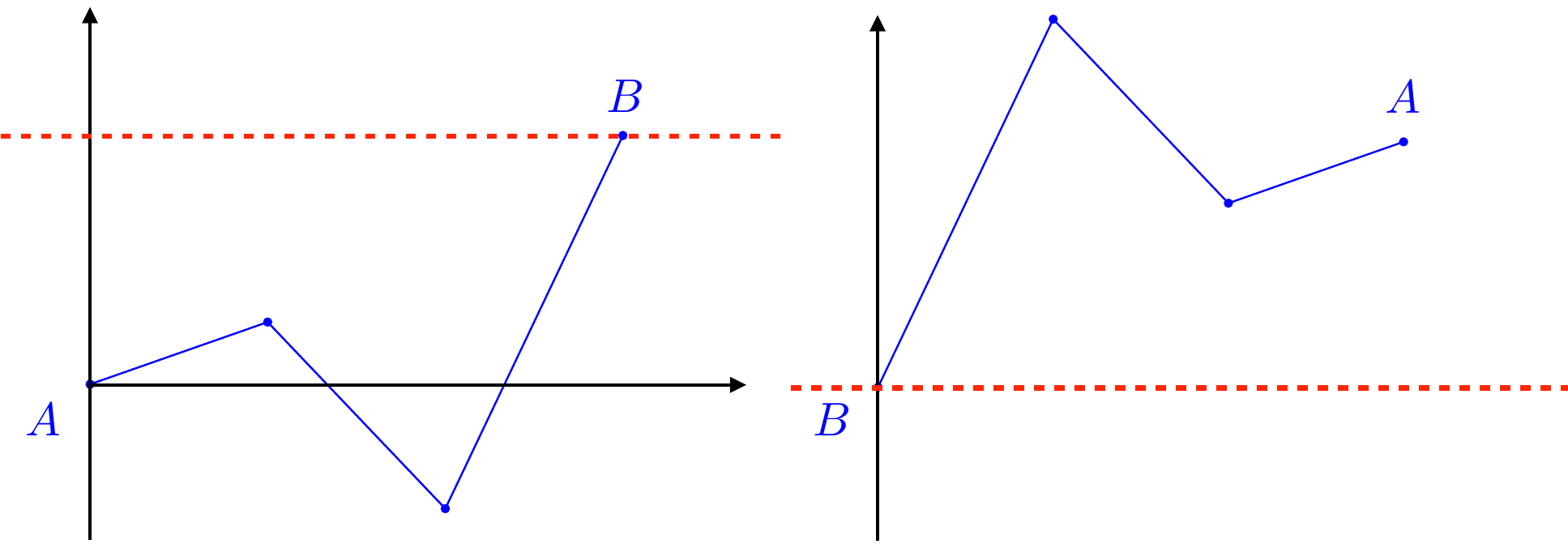}
\caption{
The dual walk on the right is obtained from the original walk on the left by fixing the endpoint
$B$ at the origin, then performing a $180^{\circ}$ rotation.}
\label{fig:duality}
\end{center}
\end{figure}

Consider the event
\be
\{\mathrm{a\;record\;occurs\;at}\; n\}
=\{x_n>x_0,\ x_n>x_1,\dots,\ x_n>x_{n-1}\},
\ee
which corresponds by duality to the event
$\{\xs_1>0,\dots,\ \xs_n>0\}$.
For example $\eta_1=1,\ \eta_2=-3,\ \eta_3=6$ generates the path $\{x_1=1,\ x_2=-2,\ x_3=4\}$.
By duality, we have
$\etas_1=6,\ \etas_2=-3,\ \etas_3=1$, which generates the path
$\{\xs_1=6,\ \xs_2=3,\ \xs_3=4\}$, as illustrated in figure~\ref{fig:duality}.

The probabilities of the following two events are thus equal~\cite{spitzer2,feller2}
\beq\label{eq:dual}
\prob(x_n>x_0,\dots,\ x_n>x_{n-1},\ x_n<x)
=\prob(x_1>0,\dots,\ x_n>0,\ x_n<x).
\eeq
This reads
\be
\prob(\CC_n(x))=\prob(\BB_n(x)),
\ee
or else
\beq
\Psi_n(x)=G_n(x)\qquad(n\ge1).
\eeq
Summing over $n\ge1$ and adding the contribution of $\mathbb{I}(x\ge0)$ for $n=0$ on both sides
results in
\beq\label{eq:PsiG}
\Psi(x)=G(x).
\eeq
The renewal function is thus equal to~\cite{spitzer2,feller2}

\begin{enumerate}

\item[1.]
the average number of records in the interval $(0,x)$ (counting the record at the origin),

\item[2.]
the average number of visits to the same interval such that $x_k>0$ for all $k=1,\dots,n$.

\end{enumerate}

By combining the above results~(\ref{eq:id1}) and~(\ref{eq:PsiG}),
we have established the fundamental identity, announced in~(\ref{eq:baxter}), namely
\beq\label{eq:fondamental}
\hat g(p)=\hat\psi(p)=\frac{1}{1-\hat f(p)},
\eeq
where
$g(x)=\psi(x)$ is the renewal density of the height process, such that $g(x)\dd x$ equals

\begin{enumerate}

\item[1.]
the average number of records in $(x,x+\dd x)$ (counting the record at the origin),

\item[2.]
the average number of visits to the same interval such that $x_k>0$ for all $k=1,\dots,n$.

\end{enumerate}

\subsection{Further results from renewal theory}
\label{sec:renew}

We consider again the renewal process formed by the successive heights $H_1,\ H_2,\dots$
Assume that $\mean{H}$ and $\mean{H^2}$ are finite
(which holds whenever $\mean{\abs{\eta}^3}$ is finite).
Define the forward recurrence length (or excess length) $E_x$ by (see~(\ref{sigmadef}))
\be
\Sigma_{R_x+1}=H_1+H_2+\cdots+H_{R_x+1}=x+E_x.
\ee
Using the classical result stating that~\cite{smith} (see Appendix~\ref{appsmith})
\beq\label{eq:smith}
\mean {\Sigma_{R_x+1}}=\mean{H}(1+\mean{R_x}),
\eeq
we conclude, using~(\ref{h1res}),~(\ref{eq:psix}) and~(\ref{eq:PsiG}) that
\beq\label{eq:overshoot}
G(x)=\frac{x+\mean{E_x}}{\sqrt{D}},
\eeq
which provides, as illustrated in figure~\ref{fig:sketchx}, a pictorial representation of $G(x)$,
the homogeneous solution of the Wiener-Hopf equation~(\ref{eq:GSpitz}), which is also the renewal
function defined in~(\ref{eq:defGn}),~(\ref{eq:defG}) and~(\ref{eq:psiH}).
Moreover, this demonstrates that $\lam(x)$, as defined in~(\ref{eq:deflam}), identifies with
$\mean{E_x}$.
Another consequence of~(\ref{eq:overshoot}) is that $G(x)$ cannot grow
faster than $x$ at large $x$, since $\mean{E_x}$ tends to a constant, as we now show.

Since $\mean{H}$ is assumed to be finite, the renewal process reaches an equilibrium as
$x\to\infty$.
Specifically, $E_x$ converges to a random variable $E$ with distribution~\cite{glrenew,cox}
\beq
f_E(y)=\frac{\dd}{\dd y}\prob(E<y)=\frac{1}{\mean{H}}\int_y^\infty\dd x\, f(x),
\label{feint}
\eeq
which implies
\be
\mean{E}=\int_0^\infty\dd y\, y f_E(y)
=\frac{{\mean{H^2}}}{2{\mean{H}}}.
\ee
As a result, asymptotically,
\beq\label{eq:homo2}
G(x)\approx\frac{x+\mean{E}}{\sqrt{D}},
\eeq
recovering the asymptotic form~(\ref{eq:homo1}), where the extrapolation length is given by
\beq\label{eq:ell}
\ell=\mean{E}=\frac{\mean{H^2}}{2\sqrt{D}},
\eeq
provided this quantity is finite~\cite{spitzer2}.
The first two moments of $H$ thus read
\beq
\mean{H}=\sqrt{D},\qquad
\mean{H^2}=2\sqrt{D}\,\ell.
\label{h12res}
\eeq
One can also derive~(\ref{eq:ell}) by a more elementary method, as follows.
The mean number of
renewals occurring in $(0,x)$ (excluding the event at the origin) is given for large $x$
by~\cite{smith,cox}
\be
\mean{R_x}\approx
\frac{x}{\mean{H}}+
\left(\frac{\mean{H^2}}{2\mean{H}^2}-1\right).
\label{rasy}
\ee
This equation, along with~(\ref{h1res}),~(\ref{eq:psix}) and~(\ref{eq:PsiG}),
again leads to~(\ref{eq:ell}).

Finally,~(\ref{feint}) reads, in Laplace space,
\beq\label{felap}
\hat f_E(p)=\frac{1-\hat f(p)}{p\mean{H}},
\eeq
which, expanding both sides into power series in $p$,
leads to the following relationship between the moments of $H$ and of $E$:
\beq
\mean{E^k}=\frac{1}{k+1}\,\frac{\mean{H^{k+1}}}{\mean{H}}.
\label{femoms}
\eeq
A systematic investigation of the moments of $H$ and $E$ is presented in section~\ref{sec:moments}.
Finally, consider the backward recurrence length $B_x$, defined by
$\Sigma_{R_x}+B_x=x$.
At equilibrium, due to time reversibility, the limiting random variable $B$ shares the same
distribution as $E$, as given by~(\ref{feint}).
In particular, the extrapolation length $\ell$ also equals $\mean{B}$.

\subsection{Generating series}
\label{sec:gen}

Thus far, quantities dependent on the discrete time $n$ have been summed over $n$ to define
`stationary' quantities such as $G(x)$, $F(x)$, $\Psi(x)$ and their derivatives.
A more general perspective consists in introducing the following generating series involving an
additional
variable $s$, conjugate to $n$.

For $s$ complex with $\abs{s}<1$, we define
\beq\label{fgserdef}
f(s,x)=\sum_{n\ge1}f_n(x)s^n,\qquad
g(s,x)=\sum_{n\ge0}g_n(x)s^n.
\eeq
This leads to the decomposition, inherited from~(\ref{eq:gxdef})
\be
g(s,x)=\delta(x)+g_\reg(s,x),\qquad
g_\reg(s,x)=\sum_{n\ge1}g_n(x)s^n.
\ee
We have in particular $f(1,x)=f(x)$, $g(1,x)=g(x)$ and $g_\reg(1,x)=g_\reg(x)$.
Equation~(\ref{eq:recurr}) translates to the inhomogeneous Wiener-Hopf integral equation
\beq\label{gmilne}
g(s,x)=\delta(x)+s\int_0^\infty\dd y\, g(s,y)\, \rho(x-y)\qquad(x\ge0),
\eeq
whereas~(\ref{eq:fondam1}) translates to
\beq\label{fgmilne}
f(s,x)=s\int_0^\infty\dd y\, g(s,y)\, \rho(x+y)\qquad (x>0).
\eeq

\subsection{Recapitulation}
\label{sec:recap}

In recapitulation, this section has provided a probabilistic interpretation of Spitzer's results,
as outlined in the introduction, along with the associated quantities introduced therein.
This approach, notably, enabled the derivation of~(\ref{eq:baxter})
(or~(\ref{eq:fondamental})), which will play a central role in the forthcoming analysis.
This derivation suffices to establish that the renewal density $g(x)$ introduced in~(\ref{eq:gxdef})
identifies with the derivative of $G(x)$, the solution of the homogeneous
equation~(\ref{eq:GSpitz})\footnote{Further developments on this topic can be found
in~\cite{spitzer1,spitzer2,spitzer3}.}.
Another derivation of this identification will be presented in section~\ref{sec:anal}.

So far, the Pollaczek-Spitzer formula~(\ref{eq:Poll-S1}) has been introduced without proof.
A derivation of this formula will be given in section~\ref{sec:anal} (see~(\ref{eq:ps})),
using an alternative analytic approach based on a class of step distributions of the
form~(\ref{rhorat}).

By inserting~(\ref{eq:Poll-S1}) into~(\ref{eq:baxter}) (derived in~(\ref{eq:fondamental})
or~(\ref{fgrat})), we obtain
\beq
\hat f(p)=1-\exp(-I(p))\qquad(\Re p>0),
\label{psone}
\eeq
with
\beq
I(p)=-\frac{p}{\pi}\int_0^\infty\frac{\dd q}{p^2+q^2}\ln(1-\til\rho(q)).
\label{ione}
\eeq
An alternative form of~(\ref{ione}), obtained through integration by parts, reads
\beq
I(p)=-\frac{1}{\pi}\int_0^\infty \dd
q\,\frac{\til\rho^{\,\prime}(q)}{1-\til\rho(q)}\,\arctan\frac{q}{p}.
\label{ialt}
\eeq
For diffusive random walks, where $1-\til\rho(q)\approx D q^2$,
it is useful to rewrite~(\ref{psone}) as
\beq
\hat f(p)=1-p\sqrt{D}\exp\left(\frac{p}{\pi}\int_0^\infty\frac{\dd
q}{p^2+q^2}\ln\frac{1-\til\rho(q)}{D q^2}\right).
\label{pstwo}
\eeq
We will use either of these forms of $\hat f(p)$, depending on the circumstances.

Another fundamental result for the forthcoming developments is the Wiener-Hopf factorisation
identity~\cite{feller2}
\beq\label{eq:fellerfactor}
(1-\til f(s,q))(1-\til f(s,-q))=1- s\til \rho(q),
\eeq
where
\beq\label{eq:chi}
\til f(s,q)
=\sum_{n\ge1}s^n\int_0^\infty\dd x\,\e^{\ii q x}f_n(x)
=\mean{s^{N}\e^{\ii q H}}
\eeq
is the Fourier transform of $f(s,x)$.
In the context of random walks and L\'evy flights,
Fourier methods are particularly well-suited,
especially when the step distribution exhibits a fat tail (see~\cite[chapter XVIII]{feller2}).
As shown there, $\til f(s,q)$ satisfies
\beq\label{eq:lemma1}
\ln\frac{1}{1-\til f(s,q)}=\sum_{n\ge1}\frac{s^n}{n}\int_{0}^\infty\dd x\,\e^{\ii q x}f_{x_n}(x),
\eeq
where $f_{x_n}(x)$ is the probability density of the position $x_n$ of the walker at time $n$.
As a consequence, $\til f(s,q)$ satisfies~\eqref{eq:fellerfactor}.
Indeed, observing that
\be
\sum_{n\ge1}\frac{s^n}{n}\int_{-\infty}^\infty\dd x\,\e^{\ii q x}f_{x_n}(x)
=\sum_{n\ge1}\frac{s^n}{n}\,\til \rho(q)^n=\ln\frac{1}{1-s\til \rho(q)},
\ee
and assuming that~(\ref{eq:lemma1}) holds, we directly obtain~(\ref{eq:fellerfactor}).
It can be easily shown that~(\ref{eq:lemma1}) implies the Pollaczek-Spitzer
formula~(\ref{eq:Poll-S1}) or~(\ref{eq:ps}) (see Appendix~\ref{appequiv}).
Let us also highlight the equivalent of~(\ref{eq:lemma1}) in Laplace space, a result of
Spitzer~\cite{spitzer3}, credited by him to Baxter:
\be
\ln\frac{1}{1-\mean{s^N \e^{-p H}}}
=\sum_{n\ge1}\frac{s^{n}}{n}\int_0^\infty\dd x\,\e^{-p x}
f_{x_n}(x).
\ee

A final comment is in order.
One might hope to use~(\ref{eq:fondam1}) or \eqref{fgmilne} to evaluate $f(x)$ for an arbitrary
step distribution $\rho(x)$.
However, apart from simple step distributions such as the symmetric exponential or the symmetric
Erlang distribution, this approach proves impracticable (see~\cite{revue}).
Similarly, extracting the asymptotic tail behaviour of $f(x)$ for $\rho(x)$ given by~\eqref{rholevy},
with $0 < \th < 2$, requires considerable effort and complex calculations
(see~\cite{mounaix1,mounaix2}).

In contrast, as shown in sections~\ref{sec:asympt},~\ref{sec:moments}, and~\ref{stable},
much greater progress in the analysis can be achieved, and with considerably simpler methods, by
primarily relying on the formula~\eqref{psone} which relates $\hat f(p)$ to the Pollaczek-Spitzer
formula~(\ref{eq:Poll-S1}), where $I(p)$ is expressed using one of the forms~\eqref{ione}
or~\eqref{ialt}, or alternatively on formula~\eqref{pstwo}, along with the Wiener-Hopf
factorisation identity~\eqref{eq:fellerfactor}.
The probabilistic interpretation presented in section~\ref{sec:outline} will also prove to be of
fundamental importance.

\section{Asymptotic tail behaviour of the distribution of the first positive position}
\label{sec:asympt}

\begin{figure}[!htbp]
\begin{center}
\includegraphics[angle=0,width=.8\linewidth,clip=true]{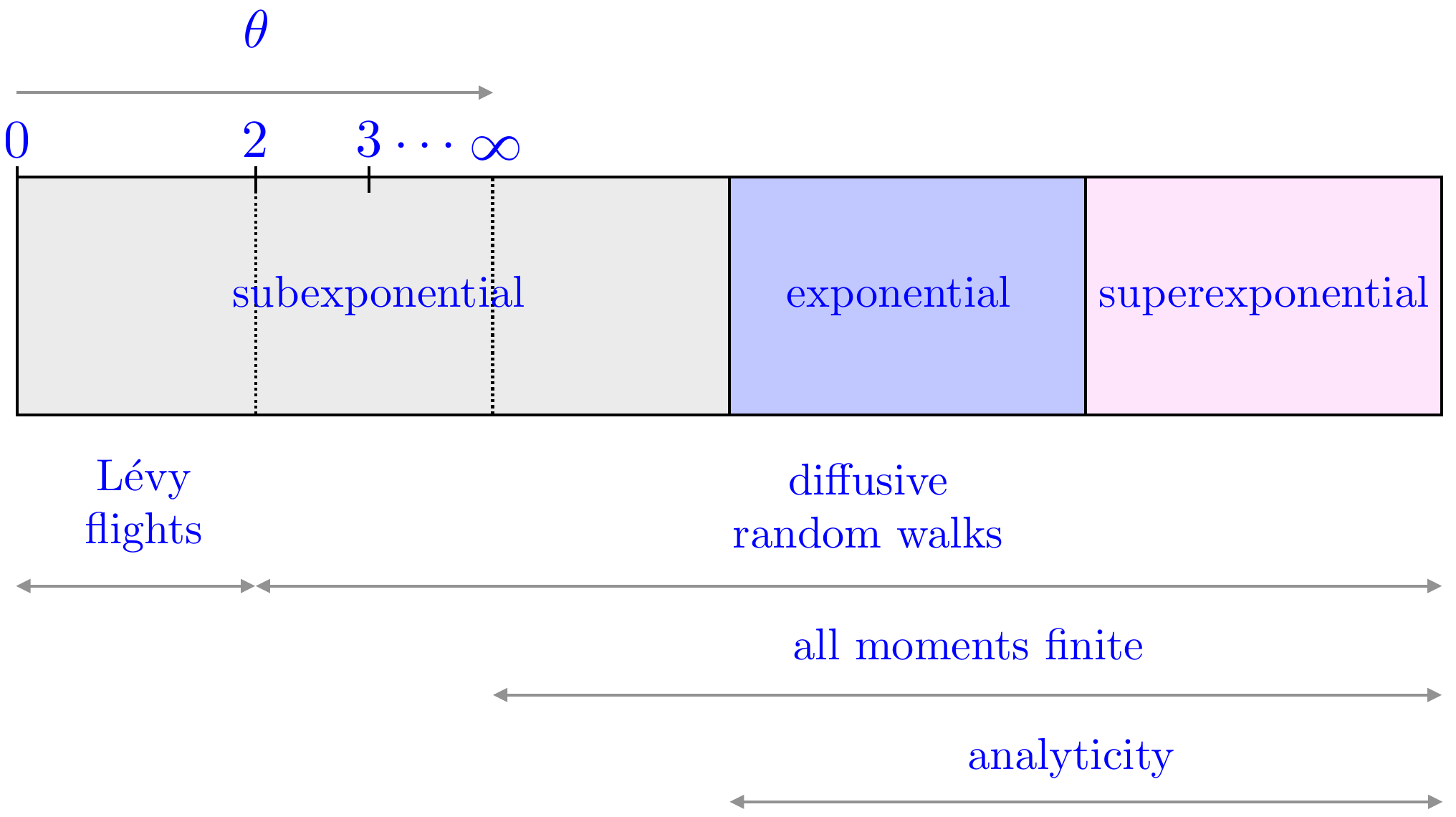}
\caption{
Schematic representation of the three classes of step distributions involved in the analysis
of the asymptotic tail behaviour of $f(x)$, in order of increasing regularity,
along with their main characteristics.}
\label{boites}
\end{center}
\end{figure}

This section is devoted to the asymptotic behaviour of the distribution $f(x)$
of the first positive position $H$ in the vicinity of its upper edge $H_\max$,
which may be either finite or infinite.
The following analysis
is entirely based on the Wiener-Hopf factorisation identity~(\ref{eq:fellerfactor}),
evaluated at $s=1$,
\beq
(1-\til f(q))(1-\til f(-q))=1-\til\rho(q),
\label{ffou}
\eeq
which holds for any continuous and symmetric step distribution $\rho(x)$.
When the latter distribution decays at least exponentially, the Fourier transforms in~(\ref{ffou})
can be continued to Laplace transforms that are analytic in a strip of the complex $p$-plane,
including the imaginary axis.
The identity~(\ref{ffou}) thus becomes
\beq
(1-\hat f(p))(1-\hat f(-p))=1-\hat\rho(p).
\label{flap}
\eeq
A systematic analysis of the asymptotic tail behaviour of $f(x)$ leads to distinguishing three classes
of step distributions, shown in different colours in figure~\ref{boites}.

\subsection{Superexponentially decaying distributions}
\label{superexp}

This first class encompasses all step distributions $\rho(x)$ for which $\hat\rho(p)$ is analytic
throughout the entire complex $p$-plane.
A first situation is where the step distribution has a finite support $(-a,a)$
(such as, e.g.,~the uniform distribution).
The distribution~$f(x)$ then has the same upper edge, i.e., $H_\max=a$.
A second situation is where the step distribution extends to infinity
and falls off more rapidly than any exponential
(such as, e.g.,~the Gaussian distribution).
The distribution $f(x)$ then also extends to infinity, i.e., $H_\max=\infty$.

Let us consider the factorisation identity~(\ref{flap}) in the $p\to-\infty$ limit.
The second factor in the left-hand side goes to unity, and so
\beq
\hat f(p)\approx\hat\rho(p)\qquad(p\to-\infty).
\eeq
This implies the asymptotic equivalence
\beq
f(x)\approx \rho(x)
\label{fas1}
\eeq
as $x$ goes to the upper edge $H_\max$ of the support of $f(x)$, whether it is finite or infinite.

\subsection{Exponentially decaying distributions}

This second class consists of the step distributions whose falloff is essentially given
by a decaying exponential, namely
\beq
\rho(x)\sim\e^{-b\abs{x}}
\label{rhoexp}
\eeq
for some $b>0$.
The symbol $\sim$ means that the leading exponential decay may be multiplied
by any prefactor having a less steep dependence on $x$, such as, e.g.,~a power of $\abs{x}$.
The Laplace transform $\hat\rho(p)$ is then analytic in the strip $\abs{\Re p\,}<b$.

Let us consider the factorisation identity~(\ref{flap}) in the $p\to-b$ limit.
The second factor in the left-hand side goes to a constant, namely $1-\hat f(b)$.
We thus obtain
\beq
\hat f(p)\approx K \hat\rho(p)\qquad(p\to-b),
\eeq
with
\beq
K=\frac{1}{1-\hat f(b)}=\hat g(b)
\label{Kdef}
\eeq
(see~(\ref{eq:baxter}),~(\ref{eq:fondamental})).
This implies the asymptotic proportionality
\beq
f(x)\approx K \rho(x)\qquad(x\to\infty).
\label{fas2}
\eeq
The proportionality constant $K$ depends on details of the step distribution.
For the symmetric exponential distribution (see section~\ref{exexp}), we have $K=2$.
For the double symmetric exponential distribution (see section~\ref{exdouble}),
\beq
K=\frac{2(p_1+p_2)}{p_1+z}
\eeq
depends on all model parameters.

\subsection{Subexponentially decaying distributions}
\label{sec:subexp}

This third class encompasses all step distributions $\rho(x)$ whose falloff is slower than any
exponential.
In such a situation, only the Fourier transform $\til \rho(q)$ is well-defined.
This situation formally amounts to taking the $b\to0$ limit of the previous one,
where the constant $K$ diverges.
This suggests that $f(x)$ should fall off (slightly) less rapidly than $\rho(x)$ at large $x$.
This heuristic expectation is corroborated by the quantitative results derived below.

\subsubsection*{Power-law decaying distributions}
\label{sec:power}

We consider first the situation where the step distribution has a power-law decay,
namely
\beq\label{rholevy}
\rho(x)\approx\frac{c}{\abs{x}^{1+\th}}\qquad(x\to\pm\infty),
\eeq
with an arbitrary tail exponent $\theta>0$.
The Fourier transform $\til \rho(q)$ of the step distribution then behaves at small $q$ as
\beq
\til \rho(q)=1-D q^2+\cdots+\til \rho_\sg(q).
\label{rhoregsg}
\eeq
The regular part consists of even powers of $q$,
while the singular part is given by
\beq
\til \rho_\sg(q)\approx2c\Gamma(-\theta)\cos(\pi\theta/2)q^\theta\qquad(q>0),
\label{rhosg}
\eeq
whenever $\theta$ is not an even integer.

For diffusive random walks, i.e., for $\theta>2$, implying a finite diffusion coefficient~$D$,
the term in $D q^2$ dominates in the expansion~(\ref{rhoregsg}).
Conversely, for superdiffusive L\'evy flights, where $\theta < 2$ and $D$ diverges, the singular
part~(\ref{rhosg}) becomes dominant.
This leads to the following classification.

\begin{enumerate}

\item[1.]
For diffusive random walks, i.e., for $\theta>2$,
the consistency of the regular and singular parts of all factors entering the
identity~(\ref{ffou}) implies
\beq
f(x)\approx\frac{a}{x^\theta}\qquad(x\to\infty).
\label{fgt}
\eeq
We have then
\beq
\til f(q)=1-\ii q\mean{H}+\cdots+a\Gamma(1-\theta)(\ii q)^{\theta-1}+\cdots
\label{fgtregsg}
\eeq
Inserting the expansions~(\ref{rhoregsg}),~(\ref{rhosg}) and~(\ref{fgtregsg}) into~(\ref{ffou})
and identifying terms, we recover $\mean{H}=\sqrt{D}$ (see~(\ref{h1res})),
and predict the amplitude
\beq
a=\frac{c}{\theta\sqrt{D}}.
\label{agt}
\eeq
So, for $\theta>2$,
the tail exponent of $f(x)$ is one unit below that of $\rho(x)$,
and the amplitudes $a$ and $c$ satisfy the linear relationship (\ref{agt}).

\item[2.]
For superdiffusive walks (L\'evy flights), i.e., for $\theta<2$,
the consistency of the leading singular parts of all factors entering the identity (\ref{ffou})
implies
\beq\label{flt}
f(x)\approx\frac{a}{x^{1+\theta/2}}\qquad(x\to\infty).
\eeq
We have then
\beq
\til f(q)=1+a\Gamma(-\theta/2)(\ii q)^{\theta/2}+\cdots
\label{fltregsg}
\eeq
Inserting the expansions~(\ref{rhoregsg}) and~(\ref{fltregsg}) in~(\ref{ffou}) yields
\beq
a=R(\theta)\sqrt{c},
\label{alt}
\eeq
with
\be
R(\theta)=\Gamma(1+\theta/2)\left(\frac{\sin(\pi\theta/2)}{\pi\Gamma(1+\theta)}\right)^{1/2}
\qquad(0<\theta<2).
\ee
So, for $\theta<2$,
the tail exponent of $f(x)$ is half that of $\rho(x)$, confirming a result of Sinai~\cite{sinai},
and the amplitudes $a$ and $c$ satisfy the nonlinear relationship~(\ref{alt}).
An equivalent result is given in~\cite{gmsprl,revue,msbook}.
The function $R(\theta)$ vanishes at both endpoints,~as
\be
R(\theta)\approx\sqrt{\frac{\theta}{2}}\quad(\theta\to0),\qquad
R(\theta)\approx\frac{\sqrt{2-\theta}}{2}\quad(\theta\to2).
\ee
Its maximum $R_\max=0.507018$ is reached for $\theta=0.857060$.
The value $R(1)=1/2$ is hardly below $R_\max$.

\item[3.]
In the marginal case where $\theta=2$,
i.e.,
\be
\rho(x)\approx\frac{c}{\abs{x}^3},
\ee
so that the diffusion coefficient $D$ is logarithmically divergent,
we mention, skipping details, that the same line of reasoning yields
a logarithmic correction of the form
\beq
f(x)\approx\frac{1}{2x^2}\left(\frac{c}{\ln x}\right)^{1/2}.
\label{fmargin}
\eeq

\end{enumerate}

\begin{figure}[!htbp]
\begin{center}
\includegraphics[angle=0,width=.7\linewidth,clip=true]{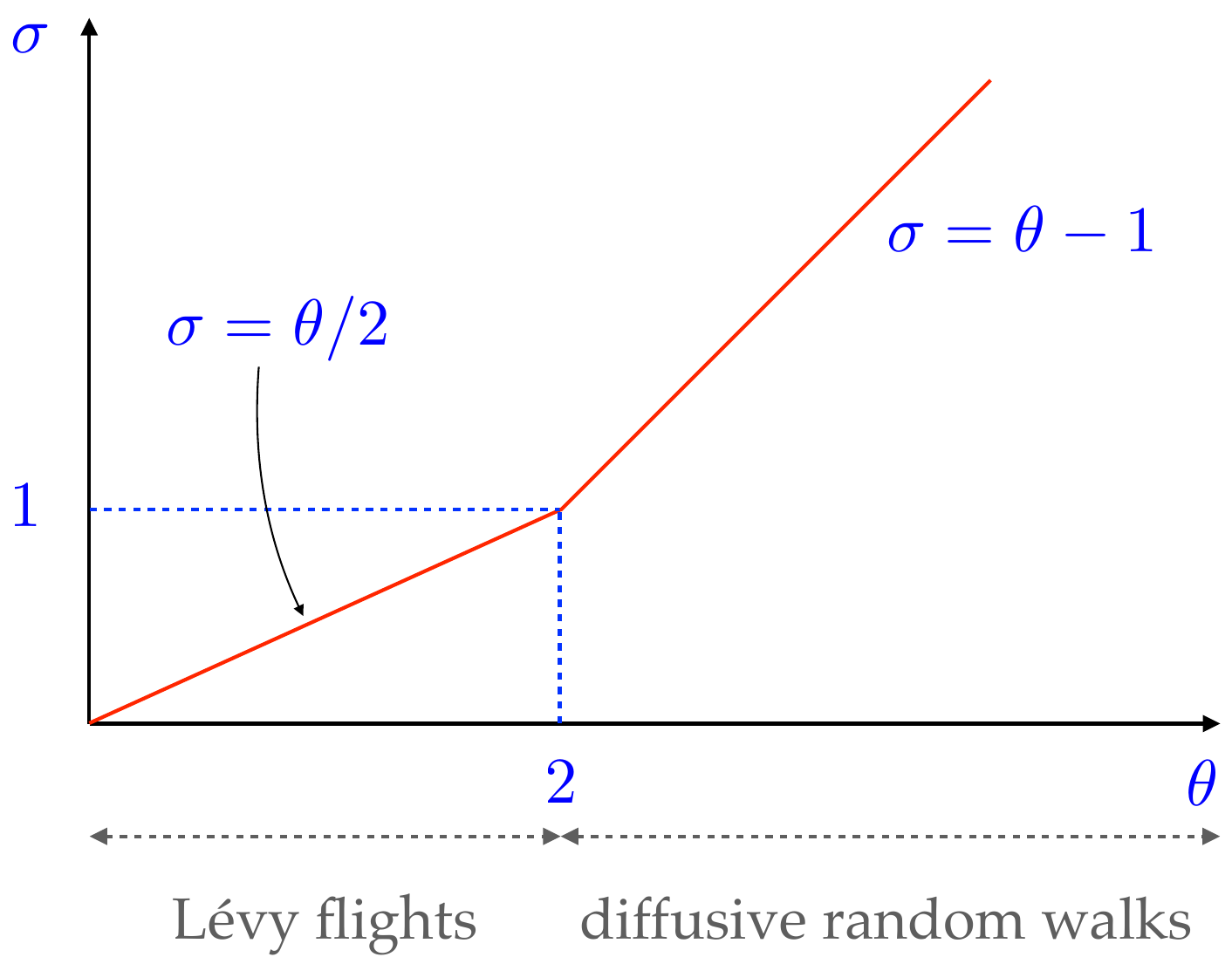}
\caption{
Dependence of the tail exponent $\sigma$
characterising the decay of the distribution $f(x)$ of the first positive position
on the tail exponent $\theta$ of the step distribution, as given by~(\ref{alphares}).}
\label{alphaplot}
\end{center}
\end{figure}

To sum up, whenever the step distribution has a power-law decay of the form~(\ref{rholevy}),
with tail exponent $\theta$,
the distribution of the first positive position also has a power-law decay,
\be
f(x)\approx\frac{a}{x^{1+\sigma}}\qquad(x\to\infty),
\ee
with a tail exponent $\sigma$ given by (see~(\ref{fgt}),~(\ref{flt}))
\beq
\sigma=\left\{
\begin{array}{cl}
\theta/2\qquad & (\theta<2),\\
\theta-1\qquad & (\theta>2),
\end{array}
\right.
\label{alphares}
\eeq
and an amplitude $a$ given by~(\ref{agt}) or~(\ref{alt}).
Figure~\ref{alphaplot} illustrates the dependence of the tail exponent $\sigma$ on $\theta$.
The inequality $\sigma<\theta$ corroborates the heuristic expectation that the distribution $f(x)$
falls off less rapidly than the step distribution $\rho(x)$.
The break point at $\theta=2$
corresponds to the logarithmic correction given by~(\ref{fmargin}).

The preceding results have the following consequences on the moments of $H$ (see
section~\ref{sec:moments}).
For superdiffusive walks, such that $D$ is divergent (i.e., $\theta<2$), the mean value of $H$
diverges,
in agreement with~(\ref{h1res}).
For diffusive walks, such as $D$ is finite (i.e., $\theta>2$), the moments $\mean{H^k}$ are finite
for $k=1,\dots,k_\max$, with
\beq
k_\max=\Int(\sigma)=\Int(\theta)-1.
\label{kmax}
\eeq

\subsubsection*{Other subexponentially decaying distributions}
\label{sec:other}

We now turn to the analysis of step distributions $\rho(x)$ whose decay, while subexponential,
is faster than that of any negative power of $\abs{x}$, so that the tail exponent $\theta$ is formally infinite.
In such a situation, only the Fourier transform $\til \rho(q)$ is well-defined,
even though all moments of~$\abs{\eta}$ are finite.

We consider first, for definiteness, the prototypical example of such distributions, the stretched exponential
\beq
\rho(x)=A\,\e^{-\abs{x}^a},\qquad A=\frac{a}{2\Gamma(1/a)},
\label{stretch}
\eeq
with arbitrary stretching exponent $0<a<1$.
The corresponding moments read
\beq
\mean{\abs{\eta}^m}=\frac{\Gamma((m+1)/a)}{\Gamma(1/a)}.
\label{othermoms}
\eeq
The Fourier transform
\beq
\til\rho(q)=A\int_{-\infty}^\infty\dd x\, \e^{\ii q x-\abs{x}^a}
\label{fouother}
\eeq
therefore has the formal power-series expansion
\beq
\til\rho(q)
=\sum_{k\ge0}\frac{(-q^2)^k}{(2k)!}\mean{\eta^{2k}}
=\sum_{k\ge0}\frac{(-q^2)^k}{(2k)!}\,\frac{\Gamma((2k+1)/a)}{\Gamma(1/a)}.
\eeq
The above expansion is only a divergent asymptotic one,
because the moments $\mean{\eta^{2k}}$ grow faster than $(2k)!$.
In other terms, $\til\rho(q)$ is indefinitely differentiable, but not analytic.
Its singular part $\til\rho_\sg(q)$ is thus expected to be smaller than any power of $q$.
In line with the usual analysis of the divergent perturbation series
met in quantum mechanics and quantum field theory (see, e.g.,~\cite{zinnpr}),
it is natural to estimate $\til\rho_\sg(q)$ as the contribution of the non-trivial complex saddle
point~$x_*$
to the integral~(\ref{fouother}), given by
\beq
\ii q=ax_*^{a-1}.
\label{col}
\eeq
Skipping the subleading prefactor, we thus obtain
\beq
\til\rho_\sg(q)\sim\exp\left(-\,(1-a)\left(\frac{a}{\abs{q}}\right)^{a/(1-a)}\right).
\label{rhosgother}
\eeq
This is indeed an exponentially small essential singularity.

Inserting~(\ref{rhoregsg}) and its counterpart for $\til f(q)$ into~(\ref{ffou}),
and identifying regular and singular parts, we obtain
\beq
\ii q\sqrt{D}\,\til f_\sg(q)\approx\til\rho_\sg(q).
\label{sgother}
\eeq
with $D=\mean{\eta^2}/2=\Gamma(3/a)/(2\Gamma(1/a))$ (see~(\ref{othermoms})).
The relationship between the tails of $f(x)$ and $\rho(x)$ can be derived,
to leading order at large $x$,
by invoking again the saddle-point approximation,
but without having to perform any explicit derivation.
In line with~(\ref{col}),
it is indeed sufficient to replace in~(\ref{sgother}) $\ii q$ by $ax^{a-1}$.
We thus obtain
\beq
f(x) \approx \frac{x^{1-a}}{a\sqrt{D}}\,\rho(x).
\label{resother}
\eeq
So, for the stretched exponential step distribution~(\ref{stretch}),
the tail of the distribution~$f(x)$ is nearly identical to that of the step distribution itself,
differing only by the prefactor given explicitly in~(\ref{resother}), involving a power of $x$.
This result matches previous ones at both endpoints of the range of the stretching exponent $a$.
As $a\to0$, disregarding amplitudes,~(\ref{resother}) involves a factor $x$,
consistent with the relation $\sigma=\theta-1$ between the two tail exponents (see~(\ref{alphares})).
As $a\to1$,~(\ref{resother}) predicts an asymptotic proportionality
between the two distributions, consistent with~(\ref{fas2}).

The above analysis can be readily extended to any step distribution with a tail of the form
\be
\rho(x)\sim\exp\left(-\varphi(x)\right),
\ee
where $\varphi(x)$ is a smoothly increasing function
whose growth at large $x$ is much faster than $\ln x$ and much slower than $x$.
In such a circumstance, the key formulas~(\ref{col}) and~(\ref{resother}) respectively generalise to
\be
\ii q=\varphi'(x^*)
\ee
and
\beq
f(x)\approx\frac{\rho(x)}{\sqrt{D}\,\varphi'(x)}.
\label{othergal}
\eeq
The derivative $\varphi'(x)$ slowly goes to zero at large $x$.
This behaviour has two consequences.
First, $x^*$ becomes large as $q\to0$, thus validating the saddle-point approach.
Second, the expression~(\ref{othergal}) corroborates our heuristic expectation
that $f(x)$ falls off slightly less rapidly than $\rho(x)$.

\section{Moments of the first positive position}
\label{sec:moments}

The purpose of this section is to determine the moments $\mean{H^k}$ of the first positive position~$H$.
We assume that the step distribution $\rho(x)$ decreases more rapidly than any power of $x$,
so that all moments of $H$ are finite.
If this is not the case, the expressions derived below are only valid for the lowest-order moments,
whose order $k$ is at most~$k_\max$, given by~(\ref{kmax}).

The analysis begins with the following observation.
A comparison between~(\ref{felap}) and~(\ref{pstwo}) yields
\beq
\hat f_E(p)=\mean{\e^{-pE}}
=\exp\left(\frac{p}{\pi}\int_0^\infty\frac{\dd q}{p^2+q^2}\ln\frac{1-\til\rho(q)}{D q^2}\right).
\label{feexp}
\eeq
Let us denote the cumulants of the excess length $E$ by $c_k$
and the corresponding generating series by
\beq
K(p)=\sum_{k\ge1}\frac{c_k}{k!}\,(-p)^k=\ln\mean{\e^{-pE}}=\ln\hat f_E(p).
\label{kserdef}
\eeq
In view of~(\ref{feexp}), we have
\beq
K(p)=\frac{p}{\pi}\int_0^\infty\frac{\dd q}{p^2+q^2}\ln\frac{1-\til\rho(q)}{D q^2},
\label{kint}
\eeq
which is related to the integral $I(p)$ defined in~(\ref{ione}) as
\beq
K(p)=-I(p)-\ln(p\sqrt{D}).
\label{kirel}
\eeq

The integral expression~(\ref{kint}) allows to determine all the cumulants $c_k$ of $E$.
This is analysed in detail below.
Even cumulants are given by the series~(\ref{evenser}),
yielding the polynomial expressions~(\ref{ceven}) in terms of even moments of the step distribution.
Odd cumulants will be given more intricate expressions involving subtracted integrals (see~\eqref{codd}).
The first of them identifies with the extrapolation length, i.e.,
\be
c_1=\mean{E}=\ell,
\ee
in agreement with~(\ref{eq:ell}).
Its expression~(\ref{codd}) agrees with~(\ref{eq:l-Milne}), as should be.

The moments of the excess length $E$ are expressed in terms of its cumulants through the Bell polynomials,
\be
\mean{E^k}=B_k(c_1,c_2,\dots,c_k),
\ee
that is, explicitly,
\beqa
\mean{E}&=&c_1,
\nonumber\\
\mean{E^2}&=&c_2+c_1^2,
\nonumber\\
\mean{E^3}&=&c_3+3c_1c_2+c_1^3,
\nonumber\\
\mean{E^4}&=&c_4+4c_1c_3+3c_2^2+6c_1^2c_2+c_1^4,
\nonumber\\
\mean{E^5}&=&c_5+5c_1c_4+10c_2c_3+10c_1^2c_3+15c_1c_2^2+10c_1^3c_2+c_1^5,
\eeqa
and so forth.

The moments of the first positive position $H$ are related to those of $E$ by~(\ref{femoms}).
We thus obtain the moments of $H$ in terms of the cumulants of $E$,
\be
\mean{H^{k+1}}=(k+1)\sqrt{D}\,B_k(c_1,c_2,\dots,c_k),
\ee
that is, explicitly\footnote{Note that $B_0=1$.},
\beqa\label{moms}
\mean{H}&=&\sqrt{D},
\nonumber\\
\mean{H^2}&=&2\sqrt{D}\,c_1,
\nonumber\\
\mean{H^3}&=&3\sqrt{D}\,(c_2+c_1^2),
\nonumber\\
\mean{H^4}&=&4\sqrt{D}\,(c_3+3c_1c_2+c_1^3),
\nonumber\\
\mean{H^5}&=&5\sqrt{D}\,(c_4+4c_1c_3+3c_2^2+6c_1^2c_2+c_1^4),
\nonumber\\
\mean{H^6}&=&6\sqrt{D}\,(c_5+5c_1c_4+10c_2c_3+10c_1^2c_3+15c_1c_2^2+10c_1^3c_2+c_1^5),
\eeqa
and so forth.

The universal expression of the first moment $\mean{H}$ has been known
for long~\cite{spitzer2,feller2} (see~(\ref{h1res})).
The expression of the second moment $\mean{H^2}$ agrees with~(\ref{eq:ell}),~(\ref{h12res}).
The variance of $H$ reads
\beq
\mean{H^2}-\mean{H}^2=(2\Az-1)D
\label{varh}
\eeq
in terms of the dimensionless quantity $\Az$ introduced in~(\ref{abdef}),
which therefore satisfies the inequality $\Az>1/2$.

We now turn to the derivation of explicit expressions for the cumulants $c_k$ of $E$.
As already said above, even and odd values of $k$ have to be dealt with separately.

\subsubsection*{Even cumulants of $E$}

The derivation of the even cumulants $c_{2m}$ relies on the Wiener-Hopf factorisation
identity~(\ref{flap}).
By substituting~(\ref{felap}) into~(\ref{flap}), we obtain:
\be
\hat f_E(p)\hat f_E(-p)=\frac{\hat\rho(p)-1}{Dp^2},
\ee
i.e.,
\beq
K(p)+K(-p)=\ln\frac{\hat\rho(p)-1}{Dp^2}.
\eeq
Using~(\ref{kserdef}), this yields
\beq
\sum_{m\ge1}\frac{c_{2m}}{(2m)!}p^{2m}=\frac12\ln\frac{\hat\rho(p)-1}{Dp^2}.
\label{evenser}
\eeq
The even cumulants follow by expanding the right-hand side, which results in
\beqa
c_2&=&\frad{\mean{\eta^4}}{24D},\qquad
c_4=\frad{\mean{\eta^6}}{60D}-\frad{\mean{\eta^4}^2}{96D^2},\qquad
c_6=\frad{\mean{\eta^8}}{112D}-\frad{\mean{\eta^4}\mean{\eta^6}}{48D^2}+\frad{5\mean{\eta^4}^3}{576D^3},
\nonumber\\
c_8&=&\frad{\mean{\eta^{10}}}{180D}-\frad{\mean{\eta^4}\mean{\eta^8}}{48D^2}-\frad{7\mean{\eta^6}^2}{360D^2}
+\frad{7\mean{\eta^4}^2\mean{\eta^6}}{144D^3}-\frad{35\mean{\eta^4}^4}{2304D^4},
\label{ceven}
\eeqa
and so forth.
We recall that $\mean{\eta^2}=2D$.

\subsubsection*{Odd cumulants of $E$}

The evaluation of the odd cumulants is more intricate.
An efficient approach consists in introducing the Mellin transform $M_K(s)$ of the function $K(p)$
defined in~(\ref{kint}),
\be
M_K(s)=\int_0^\infty\dd p\,p^{s-1}\,K(p)
=\frac{1}{\pi}\int_0^\infty\dd p\,p^s
\int_0^\infty\frac{\dd q}{p^2+q^2}\ln\frac{1-\til\rho(q)}{D q^2}.
\ee
Interchanging the order of integrations and using the identity
\be
\int_0^\infty\dd p\,\frac{p^s}{p^2+q^2}=\frac{\pi\,q^{s-1}}{2\cos(\pi s/2)}\qquad(-1<\Re s<1),
\ee
we obtain
\beq
M_K(s)=\frac{\mu_K(s)}{2\cos(\pi s/2)},
\label{mkprod}
\eeq
with
\be
\mu_K(s)=\int_0^\infty\dd q\,q^{s-1}\,\ln\frac{1-\til\rho(q)}{D q^2}\qquad(-2<\Re s<0).
\ee
The expression~(\ref{mkprod}) holds for $-1<\Re s<0$.
The inverse Mellin formula reads
\be
K(p)=\int\frac{\dd s}{2\pi\ii}\,p^{-s}\,M_K(s),
\ee
where the integral runs along a vertical contour in the strip $-1<\Re s<0$.
The power-series expansion of $K(p)$ is obtained by collecting the contributions
of all the poles of $M_K(s)$ in the left-hand half-plane ($\Re s<0$).

\begin{enumerate}

\item[1.]
Even cumulants are in correspondence with the poles of $\mu_K(s)$ at $s=-2m$ with $m=1,2,\dots$
We have\footnote{The symbol $[x^n]f(x)$ denotes the coefficient of $x^n$ in the power-series
expansion of $f(x)$.}
\beq\label{eq:coeff}
[p^{2m}]K(p)=\frac{c_{2m}}{(2m)!}=\frac{(-1)^m}{2}[q^{2m}]\ln\frac{1-\til\rho(q)}{D q^2}.
\eeq
This expression is equivalent to~(\ref{evenser}).

\item[2.]
Odd cumulants are in correspondence with the poles of $1/(2\cos(\pi s/2))$
at $s=-2m-1$ with $m=0,1,\dots$
We have
\beq
[p^{2m+1}]K(p)=-\frac{c_{2m+1}}{(2m+1)!}=\frac{(-1)^m}{\pi}\,\mu_K(-2m-1).
\label{muodd}
\eeq
The integral expression~(\ref{melsub}) for $\mu_K(-2m-1)$, whose
proof is provided in Appendix~\ref{appsub}, yields
\beq
c_{2m+1}=(-1)^{m+1}\frac{(2m+1)!}{\pi}
\int_0^\infty\frac{\dd q}{q^{2m+2}}
\left(\ln\frac{1-\til\rho(q)}{Dq^2}-2\sum_{n=1}^m\frac{c_{2n}}{(2n)!}(-q^2)^n\right).
\label{codd}
\eeq
\end{enumerate}

This integral formula generalises the expression~(\ref{eq:l-Milne}) for the extrapolation length~$\ell$.
The latter is recovered for $m=0$, the only case where no subtraction is involved.
Unlike~(\ref{ceven}), which are fully explicit,
the expressions~(\ref{eq:l-Milne}) and~(\ref{codd}) can be evaluated in closed form for only a few step distributions---essentially the Gaussian distribution (see~(\ref{elljgau}))
and the class of distributions of the form~(\ref{rhorat}) (see~(\ref{cksum})).
For generic step distributions, even the simplest ones, such as the uniform distribution,
the formulas~(\ref{eq:l-Milne}) and~(\ref{codd}) only lend themselves to numerical evaluation.

The expressions~(\ref{ceven}) and~(\ref{codd}) demonstrate that $c_k$ is finite whenever $\mean{\abs{\eta}^{k+2}}$ converges,
which essentially amounts to $\rho(x)$ falling off faster than $1/\abs{x}^{k+3}$.
In terms of the tail exponent $\theta$ (see~(\ref{rholevy})), this reads $\theta>k+2$.
In particular, we recover that the extrapolation length $\ell$ is finite for $\theta>3$.
Hence the moment $\mean{H^k}$ is finite whenever $\rho(x)$ falls off faster than $1/\abs{x}^{k+2}$,
i.e., $\theta>k+1$, in agreement with~(\ref{kmax}).

To close, we recall that the Wiener-Hopf factorisation~(\ref{ffou})
yields an infinite sequence of identities relating the moments of $\eta$ and $H$~\cite{alev}.
For a continuous symmetric step distribution these relations take the form:
\beq
\mean{\eta^{2n}}=\sum_{k=1}^{2n-1}(-1)^{k-1}{2n\choose k}\mean{H^k}\mean{H^{2n-k}},
\label{alevin}
\eeq
with the following first examples:
\bea
\mean{\eta^2}&=&2\mean{H}^2,
\nonumber\\
\mean{\eta^4}&=&8\mean{H}\mean{H^3}-6\mean{H^2}^2,
\nonumber\\
\mean{\eta^6}&=&12\mean{H}\mean{H^5}-30\mean{H^2}\mean{H^4}+20\mean{H^3}^2,
\nonumber\\
\mean{\eta^8}&=&16\mean{H}\mean{H^7}-56\mean{H^2}\mean{H^6}+112\mean{H^3}\mean{H^5}-70\mean{H^4}^2.
\eea
The expressions~(\ref{moms}) of the moments of $H$ satisfy these relations, as expected.
The key element in verifying this lies in the expressions~(\ref{ceven}) for the even cumulants of $E$.

\section{Stable distributions: Gauss, Cauchy, L\'evy}
\label{stable}

\subsection{A reminder of definitions}
L\'evy stable distributions naturally arise in the study of random walks due to the generalised
central limit theorem (see, e.g.,~\cite{levy,gnedenko,bouchaud,MK}).
This theorem states that for a sequence of iid random variables $X_1, X_2, \ldots$,
if the distribution of the sum $S_n=\sum_{i=1}^n X_i$, appropriately normalised, converges to a
limiting distribution as $n \to \infty$, then this limiting distribution is stable.
Equivalently, the distribution of $X$ is said to belong to the domain of attraction of a stable
distribution.

Among the classes of symmetric distributions considered in section~\ref{sec:asympt}, superexponentially
decaying distributions, exponentially decaying distributions, and subexponentially decaying
distributions with $\theta \geq 2$, belong to the domain of attraction of the stable law with
$\alpha=2$.
In contrast, subexponentially decaying distributions with $\theta<2$ belong to the domain of
attraction of the stable law with $\alpha=\theta$.

The distribution of the random variable $X$ is (strictly) stable\footnote{The general definition of
a stable distribution requires the existence of constants $c_n$ and $b_n$ such that
$X_1+\cdots+X_n \overset{d}{=} c_n X+b_n$.
Here we restrict ourselves to the case where the distribution of $X$ is symmetric.
A symmetric stable distribution is necessarily strictly stable~\cite{feller2}.} if there exist
constants $c_n > 0$ such that, for any $n\ge1$,
\beq\label{stabequiv}
X_1+\cdots+X_n \overset{d}{=} c_n X.
\eeq
The short-hand notation $\overset{\rm d}{=}$ indicates that the random variables on either side of
the equal sign have the same distribution.
The norming constants $c_n$ are necessarily of the form $c_n=n^{1/\alpha}$ for some
$0<\alpha\le2$~\cite{feller2}.
The constant $\alpha$ is called the \textit{index} or \textit{characteristic exponent} of the
distribution.
In Fourier space, the symmetric stable distribution of index $\alpha$ is given by
\beq\label{eq:rotildeq}
\til\rho(q)=\e^{-\abs{q}^\alpha},
\eeq
and so
\beq
\rho(x)=\frac{1}{\pi}\int_0^\infty\dd q\,\cos qx\,\e^{-q^\alpha}.
\label{stab}
\eeq
All symmetric stable distributions fall off monotonically on either side of their maximum
\beq
\rho(0)=\frac{\Gamma(1+1/\alpha)}{\pi}.
\label{rhozero}
\eeq
The Gaussian (or normal) distribution is stable with $\alpha=2$,
while the Cauchy distribution corresponds to $\alpha=1$.
For $0<\alpha<2$, the stable distribution has a fat tail with exponent $\alpha$:
$\rho(x)$ decays as a power law of the form~(\ref{rholevy}) with $\theta=\alpha$, namely
\beq
\rho(x)\approx\frac{c}{\abs{x}^{1+\alpha}},\qquad c=\frac{\Gamma(1+\alpha)\sin(\pi\alpha/2)}{\pi}.
\label{clevy}
\eeq

Let us now apply these general considerations to the problem at hand.
Hereafter, we shall consider random walks with step distributions that are stable with $\alpha=2$
(see section~\ref{sec:exgauss}),
$\alpha=1$ (see section~\ref{excauchy}),
and $0<\alpha<2$ (see section~\ref{exlevy}).
The definition~(\ref{stabequiv}) (with $X=\eta$ and $c_n=n^{1/\alpha}$) implies that the
distribution of the position $x_n$
of the walker at time $n$ is given in terms of the symmetric stable step distribution as
\be
f_{x_n}(x)=n^{-1/\alpha}\,\rho(n^{-1/\alpha}x).
\ee
Note that this suffices to determine the quantity $\omega$ defined in~(\ref{omegadef}).
Since $f_{x_n}(0)=n^{-1/\alpha}\rho(0)$, where $\rho(0)$ is given
in~(\ref{rhozero}),~(\ref{omegareturns}) yields
\beq
\omega=\frac{\Gamma(1+1/\alpha)\zeta(1+1/\alpha)}{\pi},
\label{omegastab}
\eeq
where $\zeta$ denotes the Riemann zeta function.

\subsection{Gaussian random walk}
\label{sec:exgauss}

The stable distribution~(\ref{stab}) with $\alpha=2$ is the Gaussian (or normal) distribution
\beq
\rho(x)=\frac{\e^{-x^2/4}}{2\sqrt{\pi}},
\label{rhogau}
\eeq
so that $\mean{\eta^2}=2$, i.e., $D=1$.
The Fourier (see~\eqref{eq:rotildeq}) and Laplace transforms of this density read
\be
\til \rho(q)=\e^{-q^2},\qquad \hat\rho(p)=\e^{p^2}.
\ee
Thus, according to~(\ref{psone}), $\hat f(p) = 1 - \exp(-I(p))$, with (see~\eqref{ione})
\beq
I(p)=-\frac{p}{\pi}\int_0^\infty\frac{\dd q}{p^2+q^2}\,\ln(1-\e^{-q^2}).
\label{igau}
\eeq

The Gaussian density~(\ref{rhogau}) decays superexponentially
(see section~\ref{superexp}),
which implies that $\hat\rho(p)$ and $\hat f(p)$ are entire functions,
analytic throughout the entire complex $p$-plane.
Systematic expansions of $\hat f(p)$ for $p\to0$ and $p\to+\infty$
can be obtained by using again the Mellin transformation, along the lines of
section~\ref{sec:moments}.
In the present situation,
it is preferable to consider the Mellin transform $M_I(s)$ of $I(p)$.
The main reason is that $I(p)$ and $M_I(s)$ still make sense for the L\'evy laws considered
hereafter
(see sections~\ref{excauchy} and~\ref{exlevy}),
where $K(p)$ is not defined.
The Mellin transform
\beq
M_I(s)=\int_0^\infty\dd p\,p^{s-1}\,I(p),
\eeq
reads, in analogy with~(\ref{mkprod}),
\beq
M_I(s)=-\frac{\mu_I(s)}{2\cos(\pi s/2)},
\label{miprod}
\eeq
where
\be
\mu_I(s)=\int_0^\infty\dd q\,q^{s-1}\,\ln(1-\e^{-q^2})=-\frac{\Gamma(s/2)\zeta(1+s/2)}{2}.
\ee
Substituting this expression for $\mu_I(s)$ into~(\ref{miprod}), we obtain
\beq
M_I(s)
=\frac{\Gamma(s/2)\,\zeta(1+s/2)}{4\cos(\pi s/2)}
=-\frac{(2\pi)^{1+s/2}\,\zeta(-s/2)}{4s\,\sin(\pi s/4)\,\cos(\pi s/2)}.
\label{melgau}
\eeq
Both expressions are valid for $0<\Re s<1$.
They are related through the reflection formula for the Riemann zeta function.
The inverse Mellin transform reads
\beq
I(p)=\int\frac{\dd s}{2\pi\ii}\,p^{-s}\,M_I(s),
\label{imellin}
\eeq
where the integral runs along a vertical contour in the strip $0<\Re s<1$.

The moments of $H$ can be determined by means of the power-series expansion of~$I(p)$,
which
is obtained by inserting the second expression of~(\ref{melgau}) into~(\ref{imellin}),
and summing the contributions of all the poles of the integrand to the left of the contour,
namely a double pole at $s=0$, and simple poles at $s=-2$, $s=-4n$ for $n=1,2,\dots$, and
$s=-(2n+1)$ for $n=0,1,\dots$.
We thus obtain
\be
I(p)=-\ln p+I_\even(p)+I_\odd(p),
\ee
with
\be
I_\even(p)=-\frac{p^2}{4}-\sum_{n\ge1}\frac{B_{2n}}{4n(2n)!}\,p^{4n}
=-\frac{1}{2}\ln\frac{\e^{p^2}-1}{p^2},
\ee
where the $B_{2n}$ denote the Bernoulli numbers,
and
\be
I_\odd(p)=\sum_{n\ge0}\frac{(-1)^{n+1}\zeta(n+1/2)}{(2n+1)\sin((2n+1)\pi/4)(2\pi)^{n+1/2}}\,p^{2n+1}.
\ee
In terms of the cumulants $c_k$ of $E$ (see~(\ref{kserdef}),~(\ref{kirel})), the above results translate to
\beqa
&&c_2=\frad{1}{2},\qquad
c_{2m}=\frad{(2m-1)!}{m!}\,B_m\quad(m\ \hbox{even}),\qquad
c_{2m}=0\quad(m\ \hbox{odd}\,\neq1),
\nonumber\\
&&c_{2m+1}=\frad{(-1)^{m+1}(2m)!\,\zeta(m+1/2)}{\sin((2m+1)\pi/4)(2\pi)^{m+1/2}}.
\label{elljgau}
\eeqa
In particular, the extrapolation length reads
\be
\ell=-\frac{\zeta(1/2)}{\sqrt{\pi}}\approx0.823916.
\ee
Inserting~(\ref{elljgau}) into~(\ref{moms}),
we obtain the following expressions for the moments of $H$:
\beqa\label{eq:momentgauss}
\mean{H}&=&1,
\nonumber\\
\mean{H^2}
&=&-\frac{2\zeta(1/2)}{\sqrt{\pi}}\approx1.647833,
\nonumber\\
\mean{H^3}
&=&\frac{3}{2\pi}(2\zeta(1/2)^2+\pi)\approx3.536516,
\nonumber\\
\mean{H^4}
&=&\frac{2}{\sqrt{\pi^3}}(2\zeta(3/2)-2\zeta(1/2)^3-3\pi\zeta(1/2))\approx9.057323,
\nonumber\\
\mean{H^5}
&=&\frac{5}{4\pi^2}(-16\zeta(1/2)\zeta(3/2)+4\zeta(1/2)^4+12\pi\zeta(1/2)^2+5\pi^2)\approx26.467489,
\nonumber\\
\mean{H^6}
&=&\frac{3}{\sqrt{4\pi^5}}(24\zeta(5/2)+40\zeta(1/2)^2\zeta(3/2)+20\pi\zeta(3/2)-4\zeta(1/2)^5
\nonumber\\
&-&20\pi\zeta(1/2)^3-25\pi^2\zeta(1/2))\approx85.897890,
\eeqa
and so forth.
Equivalent expressions up to the fourth moment can be found in~\cite{peres}.

The distribution $f(x)$ has a power-series expansion in $x$,
which is obtained through the asymptotic expansion of $I(p)$ as $p\to\infty$.
The latter is derived by inserting the first expression of~(\ref{melgau}) into~(\ref{imellin}),
and summing (minus) the contributions of the poles of the integrand at $s=2m+1$ for $m=0,1,\dots$
This reads
\beq
I(p)=\sum_{m\ge0}\frac{(-1)^m\,a_m}{p^{2m+1}},\qquad
a_m=\frac{\Gamma(m+1/2)\,\zeta(m+3/2)}{2\pi}.
\label{ipgau}
\eeq
Inserting this expansion into~(\ref{psone}),
we obtain
\be
\hat f(p)=\frac{\omega}{p}-\frac{\omega^2}{2p^2}+\cdots,
\ee
with
\beq
\omega=\frac{\zeta(3/2)}{2\sqrt{\pi}}\approx0.736937,
\label{omegau}
\eeq
in agreement with~(\ref{omegastab}).
The distribution of $H$ therefore reads
\be
f(x)=\omega-\frac{\omega^2}{2}\,x+\cdots\qquad(x\to0).
\label{gauser}
\ee
A few more terms will be given in~(\ref{levyser}) for an arbitrary index $\alpha$.

\begin{figure}[!htbp]
\begin{center}
\includegraphics[angle=0,width=.6\linewidth,clip=true]{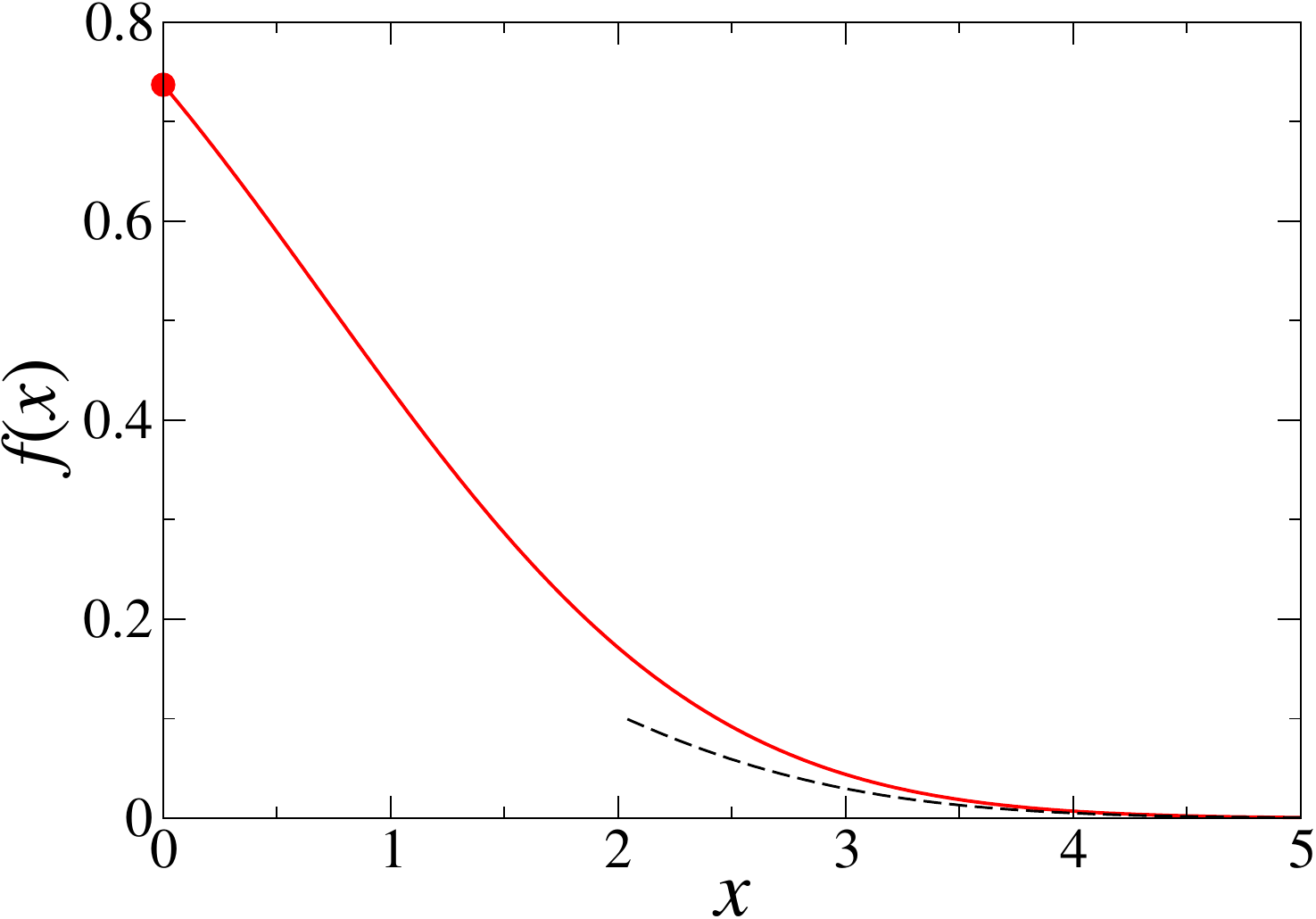}
\caption{
Distribution $f(x)$ of the first positive position for the Gaussian step
distribution~(\ref{rhogau}),
obtained by means of the numerical evaluation of 60 terms
of the power-series representation~(\ref{gauser}).
Black dashed curve: Gaussian step distribution~(\ref{rhogau}).
Symbol: $f(0)=\omega$ (see~(\ref{omegau})).}
\label{gaussplot}
\end{center}
\end{figure}

Pushing the expansion~(\ref{gauser}) to all orders
yields a convergent series representation of the distribution~$f(x)$.
The numerical evaluation of 60 terms suffices to reach an accuracy of $10^{-10}$ for $x=5$,
much better than the accuracy of standard numerical inverse Laplace techniques.
Figure~\ref{gaussplot} shows the distribution $f(x)$ thus obtained.
The black dashed curve shows the Gaussian step distribution~(\ref{rhogau}),
which provides an asymptotic equivalent to the tail of $f(x)$ at large $x$ (see~(\ref{fas1})).

\subsection{Cauchy flight}
\label{excauchy}

The stable distribution~(\ref{stab}) with $\alpha=1$ is the Cauchy distribution
\beq
\rho(x)=\frac{1}{\pi(1+x^2)}.
\label{rhocau}
\eeq
In the notation of~(\ref{rholevy}), we have $\theta=1$ and $c=1/\pi$.
The Fourier transform of~(\ref{rhocau}) is (see~\eqref{eq:rotildeq})
\be
\til \rho(q)=\e^{-\abs{q}},
\ee
so that the expression~(\ref{ialt}) of the integral $I(p)$ in~(\ref{psone}) reads
\beq
I(p)=\frac{1}{\pi}\int_0^\infty\frac{\dd q}{\e^q-1}\arctan\frac{q}{p}.
\label{ipcaudef}
\eeq
This integral appears in the second of Binet's expressions for the logarithm of Euler's
gamma function~\cite{binet,whittaker},
\be
\ln\Gamma(z)=\left(z-\frac12\right)\ln z-z+\frac12\ln(2\pi)
+2\int_0^\infty\frac{\dd t}{\e^{2\pi t}-1}\arctan\frac{t}{z}.
\ee
Setting $q=2\pi t$ and $p=2\pi z$, we obtain
\be
I(p)=\ln\Gamma\left(\frac{p}{2\pi}\right)-\frac{p}{2\pi}\left(\ln\frac{p}{2\pi}-1\right)+\frac12\ln
p-\ln(2\pi).
\ee
Inserting this expression into~(\ref{psone}),
we obtain the following remarkable closed-form expression
\beq
\hat
f(p)=1-\frac{\displaystyle{\sqrt{p}}}{\Gamma\left(\frad{p}{2\pi}+1\right)}
\left(\frac{p}{2\pi\e}\right)^{p/(2\pi)}.
\label{fcau}
\eeq
The reflection formula for the Gamma function ensures that this expression satisfies the
factorisation identity~(\ref{ffou}).

The asymptotic behaviour of the distribution $f(x)$ at large $x$ is obtained by
expanding~(\ref{fcau}) for $p\to0$ as
\be
\hat f(p)=1-\sqrt{p}+\frac{\sqrt{p^3}}{2\pi}\left(\ln\frac{2\pi}{p}+1-\gamma\right)+\cdots,
\ee
where $\gamma$ denotes Euler's constant,
which implies the following expansion:
\beq
f(x)=\frac{1}{2\sqrt{\pi x^3}}+\frac{3\ln(8\pi x)-5}{8\sqrt{\pi^3x^5}}+\cdots
\label{fcauasy}
\eeq
The leading power-law decay with exponent 3/2 agrees with the general results~(\ref{flt})
and~(\ref{alt})
with $c=1/\pi$ and $R(1)=1/2$.
Higher-order terms involve polynomials in $\ln x$ with increasing degrees.

A systematic power-series expansion of the distribution $f(x)$ can be obtained by expanding $\hat
f(p)$ for $p\to\infty$.
The integral $I(p)$, defined in~(\ref{ipcaudef}),
has the following asymptotic expansion to all orders:
\beq
I(p)=\sum_{m\ge0}\frac{B_{2m+2}}{2(m+1)(2m+1)}\left(\frac{2\pi}{p}\right)^{2m+1},
\label{ipcau}
\eeq
where $B_n$ are again the Bernoulli numbers.
This expansion is derived by expressing the arctan function in~(\ref{ipcaudef}) as a power series
and integrating term by term.
Inserting~(\ref{ipcau}) into~(\ref{psone}), we obtain
\be
\hat
f(p)=\frac{\pi}{6p}-\frac{\pi^2}{72p^2}-\frac{139\pi^3}{6480p^3}+\frac{571\pi^4}{155520p^4}+\cdots,
\ee
hence
\be
f(x)=\frac{\pi}{6}-\frac{\pi^2x}{72}-\frac{139\pi^3x^2}{12960}+\frac{571\pi^4x^3}{933120}+\cdots
\label{causer}
\ee
A few more terms will be given in~(\ref{levyser}) for an arbitrary index $\alpha$.
The value
\beq
\omega=f(0)=\frac{\pi}{6}\approx0.523598
\label{omecau}
\eeq
agrees with~(\ref{omegastab}).

\begin{figure}[!htbp]
\begin{center}
\includegraphics[angle=0,width=.6\linewidth,clip=true]{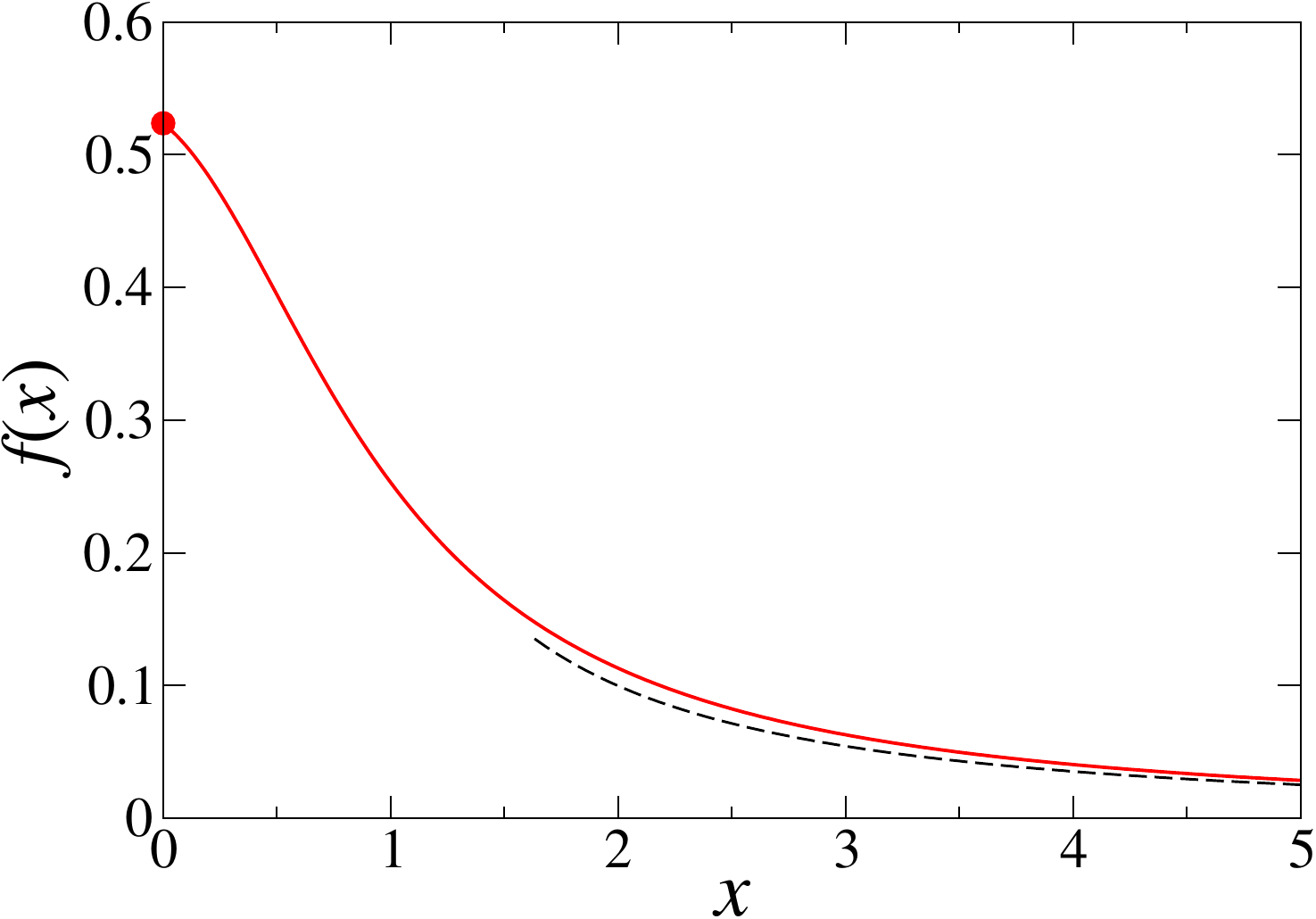}
\caption{
Distribution $f(x)$ of the first positive position for the Cauchy step distribution (\ref{rhocau}),
obtained by means of a numerical Laplace inversion of~(\ref{fcau}).
Black dashed curve: leading power-law decay given in~(\ref{fcauasy}).
Symbol: $f(0)=\omega$ (see~(\ref{omecau})).}
\label{cauchyplot}
\end{center}
\end{figure}

The expression~(\ref{fcau}) is also well-suited for a numerical Laplace inversion,
enabling thus an accurate calculation of the distribution $f(x)$ (see~figure~\ref{cauchyplot}).

\subsection{L\'evy flights}
\label{exlevy}

In this section, we examine the stable distributions~(\ref{stab}) with an arbitrary index $0 <
\alpha < 2$.
All these distributions result in superdiffusive L\'evy flights.
The forthcoming analysis closely follows that of the Gaussian distribution presented in
section~\ref{sec:exgauss}.
Many details will therefore be omitted.

In the notation of~(\ref{rholevy}), we have $\theta=\alpha$, whereas $c$ is given by~(\ref{clevy}).
Equation~(\ref{psone}) reads $\hat f(p)=1-\exp(-I(p))$, with
\beq
I(p)=-\frac{p}{\pi}\int_0^\infty\frac{\dd q}{p^2+q^2}\,\ln(1-\e^{-q^\alpha}).
\label{ilevy}
\eeq
In close analogy with~(\ref{melgau}), the Mellin transform of $I(p)$ reads
\beq
M_I(s)
=\frac{\Gamma(s/\alpha)\,\zeta(1+s/\alpha)}{2\alpha\cos(\pi s/2)}
=-\frac{(2\pi)^{1+s/\alpha}\,\zeta(-s/\alpha)}{4s\,\sin(\pi s/(2\alpha))\,\cos(\pi s/2)}.
\label{mellevy}
\eeq
Both expressions are valid for $0<\Re s<1$ and have a meromorphic continuation.

The asymptotic form of the distribution $f(x)$ at large $x$
can be obtained by expanding $I(p)$ at small $p$.
This task is carried out systematically
by substituting the second expression of~(\ref{mellevy}) into~(\ref{imellin}),
and summing the contributions of all the poles of the integrand to the left of the contour.
Assuming temporarily that the index $\alpha$ is an irrational number,
$M_I(s)$ has a double pole at $s=0$,
and simple ones at $s=-\alpha$,
at $s=-(2n+1)$ for $n=0,1,\dots$ and at $s=-2n\alpha$ for $n=1,2,\dots$
We thus obtain
\be
I(p)=-\frac{\alpha}{2}\,\ln p+\frac{p^\alpha}{4\cos(\pi\alpha/2)}+I_1(p)+I_2(p),
\ee
with
\beqa
I_1(p)&=&\sum_{n\ge0}\frac{(-1)^{n+1}\,\zeta((2n+1)/\alpha)}{(2n+1)\,(2\pi)^{(2n+1)/\alpha}\,\sin((2
n+1)\pi/(2\alpha))}\,p^{2n+1},
\nonumber\\
I_2(p)&=&-\sum_{n\ge1}\frac{B_{2n}}{4n(2n)!}\,\frac{p^{2n\alpha}}{\cos(n\pi\alpha)}.
\eeqa
Equation~(\ref{psone}) therefore reads
\beq
\hat f(p)=1-p^{\alpha/2}\exp\left(-\frac{p^\alpha}{4\cos(\pi\alpha/2)}-I_1(p)-I_2(p)\right).
\label{flevy}
\eeq
The leading singular term in $p^{\alpha/2}$ yields
\beq
f(x)\approx\frac{a}{x^{1+\alpha/2}},\qquad a=\frac{\alpha}{2\Gamma(1-\alpha/2)},
\eeq
in agreement with the general result~(\ref{alt}), where $c$ is given by~(\ref{clevy}).
Expanding the exponential in the right-hand side of~(\ref{flevy})
gives rise to terms proportional to $p^{m+n\alpha}$ for all $m,n=0,1,\dots$
Taking the first two correction terms, namely $(m,n)=(1,0)$ and $(0,1)$,
into account yields
\beqa
f(x)
&=&\frac{\alpha}{2\,\Gamma(1-\alpha/2)}\,x^{-1-\alpha/2}
\nonumber\\
&+&\frac{1}{4\,\cos(\pi\alpha/2)\,\Gamma(-3\alpha/2)}\,x^{-1-3\alpha/2}
\nonumber\\
&-&\frac{\zeta(1/\alpha)}{(2\pi)^{1/\alpha}\,\Gamma(-1-\alpha/2)\,\sin(\pi/(2\alpha))}\,x^{-2-\alpha
/2}+\cdots
\eeqa
The last two lines coalesce for $\alpha=1$, i.e., for the Cauchy distribution studied in
section~\ref{excauchy}.
This degeneracy induces the logarithmic correction term entering~(\ref{fcauasy}).
Higher-order coalescences affect higher-order correction terms whenever the index $\alpha$ is a
rational number.

The distribution $f(x)$ has a power-series expansion in $x$,
that can be obtained by means of the asymptotic expansion of $I(p)$ as $p\to\infty$.
The latter can be derived by inserting the first expression of~(\ref{mellevy}) into~(\ref{imellin}),
and summing (minus) the contributions of the poles of the integrand at $s=2m+1$ for $m=0,1,\dots$
This reads
\beq
I(p)=\sum_{m\ge0}\frac{(-1)^m\,a_m}{p^{2m+1}},\qquad
a_m=\frac{\Gamma((2m+1)/\alpha)\,\zeta(1+(2m+1)/\alpha)}{\pi\alpha}.
\label{iplevy}
\eeq
The expansions~(\ref{ipcau}) and~(\ref{ipgau}) are respectively recovered for $\alpha=1$ and
$\alpha=2$.
Inserting~(\ref{iplevy}) into~(\ref{psone}),
and performing the inverse Laplace transform term by term,
we obtain
\beqa
f(x)
&=&a_0-\frac{a_0^2}{2}\,x+\left(\frac{a_0^3}{12}-\frac{a_1}{2}\right)x^2+\left(\frac{a_0a_1}{6}
-\frac{a_0^4}{144}\right)x^3
\label{levyser}
\\
&+&\left(\frac{a_2}{24}-\frac{a_0^2a_1}{48}+\frac{a_0^5}{2880}\right)x^4
+\left(\frac{a_0^3a_1}{720}-\frac{a_0a_2}{120}-\frac{a_1^2}{240}-\frac{a_0^6}{86400}\right)x^5+\cdots
\nonumber
\eeqa
The value of $\omega=a_0$ agrees with~(\ref{omegastab}),
whereas the expansions~(\ref{causer}) and~(\ref{gauser}) are respectively recovered for $\alpha=1$ and $\alpha=2$.

To conclude, we note that extending the expansion~(\ref{levyser}) to all orders results in a convergent series representation of the distribution~$f(x)$ for $1<\alpha\le2$.

\section{A complementary analytical approach}
\label{sec:anal}

\subsection{The setting}
\label{anal:gen}

This section examines a complementary analytical approach to the distribution of the pair $(N, H)$
and, more specifically, the distribution $f(x)$ of the first positive position~$H$ reached by a
walker starting from the origin.
This approach deals with the class of step distributions $\rho(x)$
such that the bilateral Laplace transform $\hat\rho(p)$ (see~(\ref{eq:bil}))
is a rational function of $p$.
The observation that Wiener-Hopf linear integral equations, such as~(\ref{eq:GSpitz}) or~(\ref{gmilne}),
are more easily solvable for such distributions
dates back at least to the works of Wick~\cite{wick} and Chandrasekhar~\cite{chandra}
(see~\cite{chandrabook} for a review).
For this class of step distributions, most of the general results presented earlier will be
rederived in a self-contained manner.
These results will be recast into a form which strongly suggests their validity for arbitrary step
distributions.

The step distributions considered here are finite superpositions of symmetric exponentials,
of the form
\beq
\rho(x)=\frac{1}{2} \sum_{a} w_a p_a \e^{-p_a \abs{x}},
\label{rhorat}
\eeq
where the index $a$ runs over $a=1, \dots, M$.
The decay rates are assumed to be distinct and ordered as $0<p_1<\dots<p_M$.
The normalisation of $\rho(x)$ imposes the sum rule
\be
\sum_aw_a=1.
\ee
In the case where all the weights $w_a$ are positive,
the distribution~(\ref{rhorat}) is a mixture, i.e., a convex combination, of symmetric exponential
(or Laplace) distributions.
In general, the weights $w_a$ are real, albeit not necessarily positive.
They only satisfy the constraint that the density~(\ref{rhorat}) remains positive for all values of $x$.
This point will be illustrated in detail for the case $M=2$ in section~\ref{exdouble}.
As stated above, the key property of the step distribution~(\ref{rhorat}) is that its bilateral
Laplace transform is a rational function of $p$:
\be
\hat\rho(p)
=\int_{-\infty}^\infty\dd x\, \e^{-p x} \rho(x)
=\sum_a\frac{w_a p_a^2}{p_a^2-p^2}.
\ee
This expression is valid for $\abs{\Re p\,}<p_1$,
where $p_1$ is the smallest decay rate appearing in~(\ref{rhorat}).
We introduce for further reference the quantity
\beq\label{phidef}
\phi(s,p)=1-s\hat\rho(p),
\eeq
for $s$ complex with $\abs{s}<1$.
The expression~(\ref{phidef}) is an even rational function of $p$ going to unity as $\abs{p}\to\infty$.
It can therefore be factorised over its $2M$ poles $\pm p_a$ and its $2M$ zeros $\pm z_b$ as
\beq\label{phiprod}
\phi(s,p)=\frac{\prod_b(p^2-z_b^2)}{\prod_a(p^2-p_a^2)}.
\eeq
The $M$ poles $p_a$ are real and positive, and independent of $s$,
as they coincide with the decay rates appearing in~(\ref{rhorat}).
The zeros $\pm z_b$ depend on $s$ and are therefore not real in general.
It can be checked that no zero can sit on the imaginary axis.
One has indeed $\abs{1-\phi(s,\ii q)}=\abs{s}\abs{\til\rho(q)}<1$,
since each factor is less than unity in modulus.
We denote by $z_b$ the $M$ zeros of $\phi(s,p)$ with positive real parts.

\subsection{Solutions of integral equations and factorisation formulas}

For the step distributions of the form~(\ref{rhorat}),
the solutions to the key equations~(\ref{gmilne}) and~(\ref{fgmilne}) can be derived by elementary means.
It is shown in full detail in Appendix~\ref{appg} and Appendix~\ref{appf}
that the Laplace transforms
\beq
\hat g(s,p)=\int_0^\infty\dd x\, \e^{-p x}g(s,x),
\qquad
\hat f(s,p)=\int_0^\infty\dd x\, \e^{-p x}f(s,x)
\label{glapdef}
\eeq
of the solutions to~(\ref{gmilne}) and~(\ref{fgmilne})
are respectively given by the product formula
\beq
\hat g(s,p)=\frac{\prod_a(p+p_a)}{\prod_b(p+z_b)}
\label{gprod}
\eeq
and by
\beq
\hat f(s,p)=1-\frac{1}{\hat g(s,p)}.
\label{fprod}
\eeq
The result~(\ref{eq:baxter}) (or~(\ref{eq:fondamental})) is recovered for $s=1$.

The expressions~(\ref{gprod}) and~(\ref{fprod})
imply that the quantities $\hat g(s,p)$ and $\hat f(s,p)$
satisfy the factorisation formulas
\beq
\hat g(s,p)\hat g(s,-p)=\frac{1}{\phi(s,p)}
\label{ggprod}
\eeq
and
\beq
(1-\hat f(s,p))(1-\hat f(s,-p))=\phi(s,p),
\label{ffprod}
\eeq
that are central to the Wiener-Hopf approach.
Equation~(\ref{ffprod}) is the equivalent of the
Wiener-Hopf factorisation identity~(\ref{eq:fellerfactor}) in Laplace space.
Setting $s=1$ in~(\ref{ffprod}), we recover~(\ref{flap}).

\subsection{Connection with the Pollaczek-Spitzer formula}

The formula~(\ref{gprod}) obtained for $\hat g(s,p)$ can be rewritten as
\be
\hat g(s,p)
=\exp\left(\sum_a\ln(p+p_a)-\sum_b\ln(p+z_b)\right).
\ee
The product formula~(\ref{phiprod}) for $\phi(s,p)$ implies that
its logarithmic derivative has the partial fraction expansion
\be
\frac{\phi'(s,p)}{\phi(s,p)}
=\sum_b\left(\frac{1}{p+z_b}+\frac{1}{p-z_b}\right)
-\sum_a\left(\frac{1}{p+p_a}+\frac{1}{p-p_a}\right),
\label{partfrac}
\ee
with simple poles with residue $+1$ at all zeros $\pm z_b$ of $\phi(s,p)$,
and simple poles with residue $-1$ at all its poles $\pm p_a$.
We have therefore
\beq
\hat g(s,p)
=\exp\left(\int\frac{\dd r}{2\pi\ii}\frac{\phi'(s,r)}{\phi(s,r)}\ln(p+r)\right).
\label{ps2}
\eeq
Equation~(\ref{ps2}) holds (at least) for $\Re p>0$ and $0<\Re r<p_1$,
where $p_1$ is the smallest decay rate in~(\ref{rhorat}).
Using~(\ref{phidef}) and integration by parts, this equation becomes
\beqa
\hat g(s,p)&=&\exp\left(-\int\frac{\dd r}{2\pi\ii}\frac{\ln(1-s\hat\rho(r))}{p+r}\right)
\nonumber\\
&=&\exp\left(-\frac{p}{\pi}\int_0^\infty\frac{\dd q}{p^2+q^2}\ln(1-s\til \rho(q))\right).
\label{eq:ps}
\eeqa
The second expression is obtained by moving the integration contour to the imaginary axis, setting
$r=\ii q$,
and using the fact that the Fourier transform~(\ref{eq:four}), namely
\be
\til \rho(q)=\hat\rho(\ii q),
\ee
is an even real function of $q$.
The expression~(\ref{eq:ps}) identifies with the standard integral form
of the Pollaczek-Spitzer formula~\cite{spitzer,pollaczek} (see also~\cite{spitzer1,spitzer2}),
which holds for any continuous symmetric step distribution $\rho(x)$, irrespective of its decay.
For $s=1$,~(\ref{eq:ps}) reproduces~(\ref{eq:Poll-S1}).

\subsection{Connection with Sparre Andersen theory}
\label{sec:sparre}

Within the present formalism,
the probabilities $f_n$ and $g_n$ introduced in section~\ref{sec:outline}
can be evaluated by setting $p=0$ in~(\ref{fprod}) and~(\ref{ggprod}).
Given that $\hat\rho(0)=1$, it follows that $\phi(s,0)=1-s$, which leads to
\beq\label{eq:sparre}
\hat g(s,0)=\sum_{n\ge0}g_n s^n=\frac{1}{\sqrt{1-s}},\qquad
\hat f(s,0)=\sum_{n\ge1}f_n s^n=1-\sqrt{1-s}.
\eeq
Hence
\beq
g_n=b_n,\qquad
f_n=b_{n-1}-b_n=\frac{b_n}{2n-1},
\label{fgsparre}
\eeq
where $b_n$ is the binomial probability
\beq
b_n=\frac{(2n)!}{(2^n n!)^2}=\frac{{2n\choose n}}{2^{2n}}.
\label{bdef}
\eeq
Thus
\beqa
g_0&=&1,\quad
g_1=\frac{1}{2},\quad
g_2=\frac{3}{8},\quad
g_3=\frac{5}{16},\quad
g_4=\frac{35}{128},
\nonumber\\
f_1&=&\frac{1}{2},\quad
f_2=\frac{1}{8},\quad
f_3=\frac{1}{16},\quad
f_4=\frac{5}{128},
\eeqa
and so on.
At large times, we have
\be
g_n\approx\frac{1}{\sqrt{\pi n}},\qquad
f_n\approx\frac{1}{2\sqrt{\pi n^3}}.
\ee
These results are part of Sparre Andersen theory~\cite{sparre53,sparre54}.
They are universal,
in the sense that they hold irrespective of the step distribution, provided it is symmetric and
continuous,
regardless of whether we are dealing with diffusive random walks ($D$ finite)
or L\'evy flights ($D$ infinite).

\subsection{General results on the distribution of $H$}
\label{sec:fH}

The general results concerning $f(x)$ that have been derived in section~\ref{sec:outline}
can be recovered within the present formalism by taking the $s\to1$ limit.
First of all, setting $s=1$ in~(\ref{fprod}), we recover~(\ref{eq:baxter})
or~(\ref{eq:fondamental}), that is,
\be
\hat f(p)=1-\frac{1}{\hat g(p)},
\label{fgrat}
\ee
where $\hat f(p)=\hat f(1,p)$, $\hat g(p)=\hat g(1,p)$.
We have
\be
\phi(p)=\phi(1,p)=1-\hat\rho(p),
\ee
and so the factorisation formulas~(\ref{ggprod}) and~(\ref{ffprod}) read
\beq
\hat g(p)\hat g(-p)=\frac{1}{1-\hat\rho(p)},
\label{gg1prod}
\eeq
and
\beq
(1-\hat f(p))(1-\hat f(-p))=1-\hat\rho(p),
\label{ff1prod}
\eeq
which is~(\ref{flap}).

The peculiar nature of the $s\to1$ limit becomes more apparent
in the case of the step distributions of the form~(\ref{rhorat}).
The difference $\phi(p)=1-\hat\rho(p)$ vanishes for $p=0$.
In view of~(\ref{phiprod}), this implies that one of the zeros $z_b$, say $z_M$, goes to zero as
$s\to1$,
whereas the other zeros approach fixed positions with positive real parts, still denoted by $z_b$,
for $b=1,\dots,M-1$.
The expression~(\ref{phiprod}) thus becomes
\be
\phi(p)=1-\hat\rho(p)=-p^2\frac{\prod'_b(z_b^2-p^2)}{\prod_a(p_a^2-p^2)}.
\ee
Here and throughout the following, accents denote sums and products over the $M-1$ remaining zeros
$z_b$.
The diffusion coefficient reads
\beq
D=-\lim_{p\to0}\frac{\phi(p)}{p^2}=\frac{\prod'_b z_b^2}{\prod_a p_a^2}.
\label{dprod}
\eeq
The expressions~(\ref{gprod}) and~(\ref{fprod}) respectively become
\beq
\hat g(p)=\frac{\prod_a(p+p_a)}{p\,\prod'_b(p+z_b)}
\label{g1prod}
\eeq
and
\beq
\hat f(p)=1-\frac{p\,\prod'_b(p+z_b)}{\prod_a(p+p_a)}.
\label{f1prod}
\eeq
These formulas show that the distribution $f(x)$ of the first positive position $H$
has exactly the same spectrum of decay rates $(p_a)$ as the step distribution $\rho(x)$ itself,
whereas $g(x)$ is characterised by an entirely different spectrum of decay rates $(z_b)$.

\subsection{Connection with the solution of equation~(\ref{eq:GSpitz})}
\label{sec:GChandra}

Within the present framework,
the solution $G(x)$ of the homogeneous Wiener-Hopf equation~(\ref{eq:GSpitz}),
repeated here for convenience,
\be
G(x)=\int_0^\infty\dd y\,G(y)\,\rho(x-y)\qquad(x>0),
\ee
with $G(0)=1$, can be obtained along the lines of Appendix~\ref{appg}.
The Laplace transform $\hat G(p)$ of $G(x)$ reads
\beq\label{hprod}
\hat G(p)=\frac{\prod_a(p+p_a)}{p^2\,\prod'_b(p+z_b)}=\frac{\hat g(p)}{p},
\eeq
which implies
\be
g_\reg(x)=\frac{\dd}{\dd x}G(x)\qquad(x>0),
\ee
as it should (see~(\ref{eq:gxdef})).

Expanding~(\ref{hprod}) for small $p$, we obtain
\be
\hat G(p)=\frac{1}{\sqrt{D}}\left(\frac{1}{p^2}+\frac{\ell}{p}+\cdots\right),
\ee
which implies the asymptotic behaviour
\be
G(x)\approx\frac{x+\ell}{\sqrt{D}},
\ee
in agreement with~(\ref{eq:homo1}) and~(\ref{eq:homo2}),
and where the extrapolation length $\ell$ is given by
\beq
\ell=\sum_a\frac{1}{p_a}-{\sum_b}'\frac{1}{z_b}.
\label{ellsum}
\eeq
Using the identity~(\ref{partfrac}) for $s=1$, this expression can be recast as
\beq
\ell
=\int\frac{\dd p}{2\pi\ii p}\,\frac{\phi'(p)}{\phi(p)}
=-\frac{1}{\pi}\int_0^\infty\frac{\dd q}{q^2}\ln\frac{1-\til\rho(q)}{D q^2},
\label{ellint}
\eeq
thus recovering~(\ref{eq:l-Milne}).

Another quantity of interest is $\omega=f(0)=g_\reg(0)=G'(0)$, introduced in~(\ref{omegadef}).
It can be derived by expanding~(\ref{f1prod}) as $p\to+\infty$, obtaining thus $\hat
f(p)\approx\omega/p$, with
\beq
\omega=\sum_a p_a-{\sum_b}'z_b.
\label{omegasum}
\eeq
Along the lines of the derivation of~(\ref{ellint}), this can be recast as
\beq
\omega
=\int\frac{\dd p}{2\pi\ii}\,\frac{p\,\phi'(p)}{\phi(p)}
=-\frac{1}{\pi}\int_0^\infty\dd q\,\ln(1-\til\rho(q)),
\eeq
which is~(\ref{omegaint}).

\subsection{Moments of the first positive position}
\label{analmoms}

The following analysis of the moments of $H$ and $E$ parallels that exposed in
section~\ref{sec:moments}.
Within the present setting, using~(\ref{felap}),
with
\beq
\mean{H}=\sqrt{D}=\frac{\prod'_b z_b}{\prod_a p_a},
\eeq
(see~(\ref{dprod})),
the expression~(\ref{f1prod}) yields
\beq
\hat f_E(p)=\frac{\prod'_b(1+p/z_b)}{\prod_a(1+p/p_a)}.
\label{feprod}
\eeq
The generating function $K(p)$ of the cumulants $c_k$ of $E$, introduced in~(\ref{kserdef}), therefore
reads
\beq
K(p)=\ln\hat f_E(p)={\sum_b}'\ln(1+p/z_b)-\sum_a\ln(1+p/p_a).
\eeq
Expanding this result as a power series in $p$,
we obtain the following expression for the cumulants:
\beq
c_k=(k-1)!\left(\sum_a\frac{1}{p_a^k}-{\sum_b}'\frac{1}{z_b^k}\right).
\label{cksum}
\eeq
Along the lines of the derivation of~(\ref{ellint}),
this can be recast as
\beq
c_k=(k-1)!\int\frac{\dd p}{2\pi\ii\,p^k}\,\frac{\phi'(p)}{\phi(p)}.
\label{ckint}
\eeq
This expression provides an alternative starting point to derive~(\ref{codd}).

In what follows we investigate several examples of step distributions of the form~(\ref{rhorat})
where the distribution of $H$ can be investigated in some more detail.

\subsection{Symmetric exponential distribution}
\label{exexp}

The symmetric exponential step distribution,
\beq
\rho(x)=\frac{\e^{-\abs{x}}}{2},
\label{rho1}
\eeq
also known as the Laplace distribution,
corresponds to the case $M=1$ in~(\ref{rhorat}), with $p_1=1$, and hence $D=1$.
We have
\be
\phi(p)=1-\hat\rho(p)=-\frac{p^2}{1-p^2},\qquad
\hat f(p)=\frac{1}{p+1},\qquad\hat g(p)=\frac{p+1}{p},
\ee
thus
\beq
f(x)=f_E(x)=\e^{-x},\qquad g_\reg(x)=1.
\label{f1}
\eeq
The exponential distribution of $H$ has been known for long~\cite{spitzer2,feller2}
(see also~\cite{gmsprl,revue,msbook}).
In this very simple case, we have $\Az=\Bz=1$,
as well as $\mean{H^k}=\mean{E^k}=k!$ and $c_k=(k-1)!$.

The symmetric exponential step distribution
is the single example where $n$-dependent quantities will be worked out explicitly.
We have
\be
\phi(s,p)=\frac{1-s-p^2}{1-p^2},
\ee
hence $z_1=\sqrt{1-s}$.
The expressions~(\ref{gprod}) and~(\ref{fprod}) read
\be
\hat g(s,p)=\frac{p+1}{p+\sqrt{1-s}},\qquad
\hat f(s,p)=\frac{1-\sqrt{1-s}}{p+1},
\ee
so
\beqa
g_\reg(s,x)&=&(1-\sqrt{1-s})\e^{-\sqrt{1-s}\,x},\qquad
\label{gexp}
\\
f(s,x)&=&(1-\sqrt{1-s})\e^{-x}.
\label{fexp}
\eeqa

The expression~(\ref{fexp}) implies that the density $f_n(x)$ assumes a factorised form:
\be
f_n(x)=f_n\,\e^{-x}.
\label{fnexp}
\ee
In other words, the random variables $N$ and $H$ are statistically independent.
The first-passage probability $f_n$ is given by~(\ref{fgsparre}),
whereas the distribution of $H$ is exponential (see~(\ref{f1})).

The expression~(\ref{gexp}) implies that the density $g_n(x)$ is of the form
\be
g_n(x)=P_n(x)\,\e^{-x}\qquad(n\ge1),
\ee
where $P_n(x)$ is a polynomial of degree $n-1$.
We have
\beqa
P_1(x)&=&\frac{1}{2},\qquad
P_2(x)=\frac{1}{8}(2x+1),\qquad
P_3(x)=\frac{1}{16}(x+1)^2,
\nonumber\\
P_4(x)&=&\frac{1}{384}(4x^3+18x^2+30x+15),
\eeqa
and so forth.
Omitting details, simple algebra yields the general form
\be
P_n(x)=\frac{1}{2^{2n-1}}\sum_{m=0}^{n-1}T_{n-1,n-1-m}\,\frac{(2x)^m}{m!},
\ee
where the integers
\be
T_{m,k}=\frac{m+1-k}{m+1}\,{m+k\choose m}\qquad(k=0,\dots,m)
\ee
are known as the ballot numbers,
and listed as sequence number A009766 in the OEIS~\cite{OEIS}.
We have in particular
\be
P_n(0)=\frac{T_{n-1,n-1}}{2^{2n-1}}=f_n,
\ee
in agreement with~(\ref{fgnzero}) and~(\ref{fnexp}).
Finally, setting $s=1$ in~(\ref{gexp}) yields
\be
g_\reg(x)=\sum_{n\ge1}g_n(x)=1.
\ee
This identity implies the sum rule
\beq
\sum_{n\ge1}P_n(x)=\e^x.
\eeq

\subsection{Double symmetric exponential distribution}
\label{exdouble}

The next example of distributions of the form~(\ref{rhorat}) corresponds to $M=2$.
In full generality, this double symmetric exponential step distribution reads
\beq
\rho(x)=\frac12(w_1p_1\e^{-p_1\abs{x}}+w_2p_2\e^{-p_2\abs{x}})
\label{rhotwo}
\eeq
and involves three parameters.
The decay rates are assumed to be distinct and ordered as $0<p_1<p_2$.
We have
\beq
\phi(p)=1-\hat\rho(p)=-\frac{p^2(z^2-p^2)}{(p_1^2-p^2)(p_2^2-p^2)},
\eeq
where the zero $z$ is real and given by
\beq
z=\sqrt{w_1p_2^2+w_2p_1^2}.
\eeq

\begin{figure}[!htbp]
\begin{center}
\includegraphics[angle=0,width=.6\linewidth,clip=true]{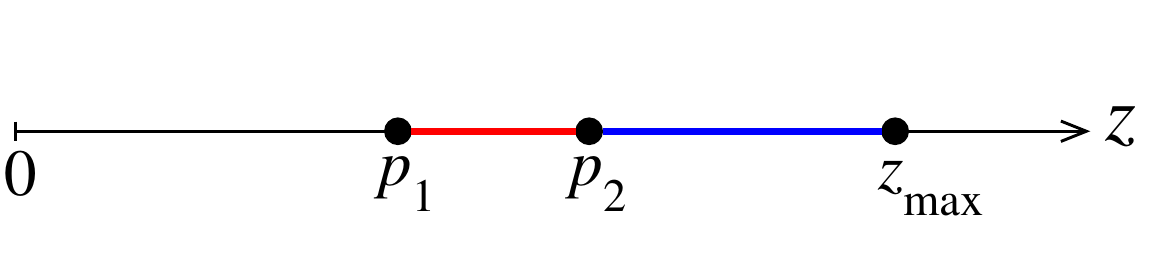}
\caption{
Domain of allowed values of the real zero $z$.
Red region ($p_1<z<p_2$): both $w_1$ and $w_2$ are positive.
Blue region ($p_2<z<z_\max$): $\rho(x)$ is positive, even though $w_2$ is negative.
At $z=z_\max$ (see~(\ref{zmaxdef})), $\rho(0)$ vanishes.}
\label{twolengths}
\end{center}
\end{figure}

It is advantageous to use $z$ as the third independent parameter, besides $p_1$ and $p_2$.
In terms of $z$, we have
\beq
w_1=\frac{z^2-p_1^2}{p_2^2-p_1^2},\qquad
w_2=\frac{p_2^2-z^2}{p_2^2-p_1^2},\qquad
\sqrt{D}=\frac{z}{p_1p_2}.
\eeq
Figure~\ref{twolengths} shows the domain over which $z$ can be varied.
For $p_1<z<p_2$ (red region), both weights $w_1$ and $w_2$ are positive.
For $p_2<z<z_\max$ (blue region), with
\beq
z_\max=\sqrt{p_1^2+p_1p_2+p_2^2}\,,
\label{zmaxdef}
\eeq
the step distribution is still positive, even though $w_2$ is negative.
For $z=z_\max$, the step distribution reads
\beq
\rho(x)=\frac{p_1p_2}{2(p_2-p_1)}(\e^{-p_1\abs{x}}-\e^{-p_2\abs{x}})
\eeq
and vanishes for $x=0$.

The general result~(\ref{f1prod}) reads
\beq
\hat f(p)=1-\frac{p(p+z)}{(p+p_1)(p+p_2)},
\eeq
yielding the following explicit expression for the distribution of $H$:
\beq
f(x)=\frac{p_1(z-p_1)\e^{-p_1x}-p_2(z-p_2)\e^{-p_2x}}{p_2-p_1}.
\label{ftwo}
\eeq

\begin{figure}[!htbp]
\begin{center}
\includegraphics[angle=0,width=.6\linewidth,clip=true]{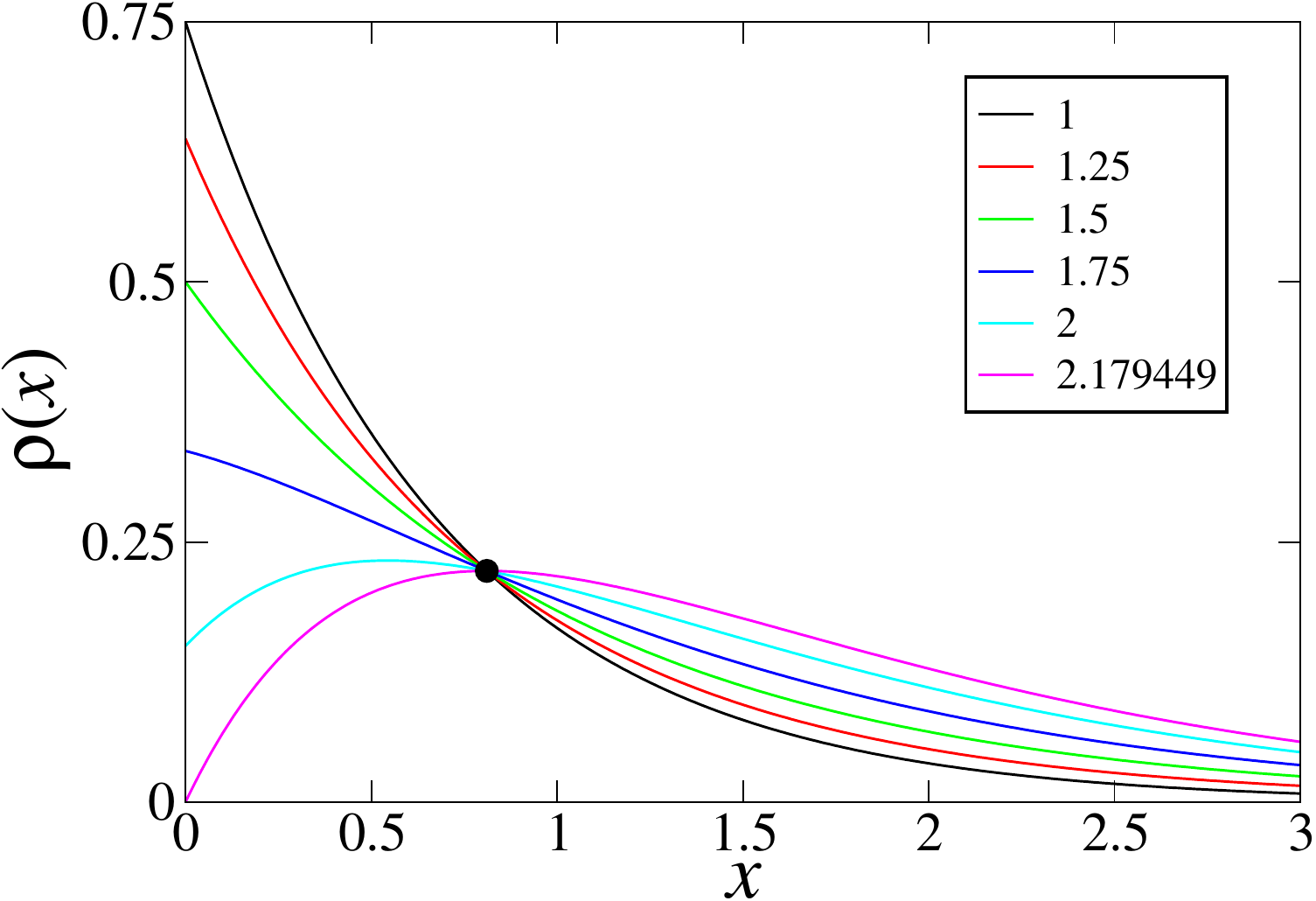}
\caption{
Positive part of the step distribution $\rho(x)$, as given by~(\ref{rhotwo}),
for $p_1=1$, $p_2=3/2$ and several $z$ (see legend),
including $z_\max=\sqrt{19}/2\approx2.179449$,
for which $\rho(0)=0$.
Black symbol: point $x_\star=2\ln(3/2)\approx0.810930$ where all curves cross
at $\rho(x_\star)=2/9$ (see~(\ref{xfstar})).}
\label{rhotwoplot}
\end{center}
\end{figure}

\begin{figure}[!htbp]
\begin{center}
\includegraphics[angle=0,width=.6\linewidth,clip=true]{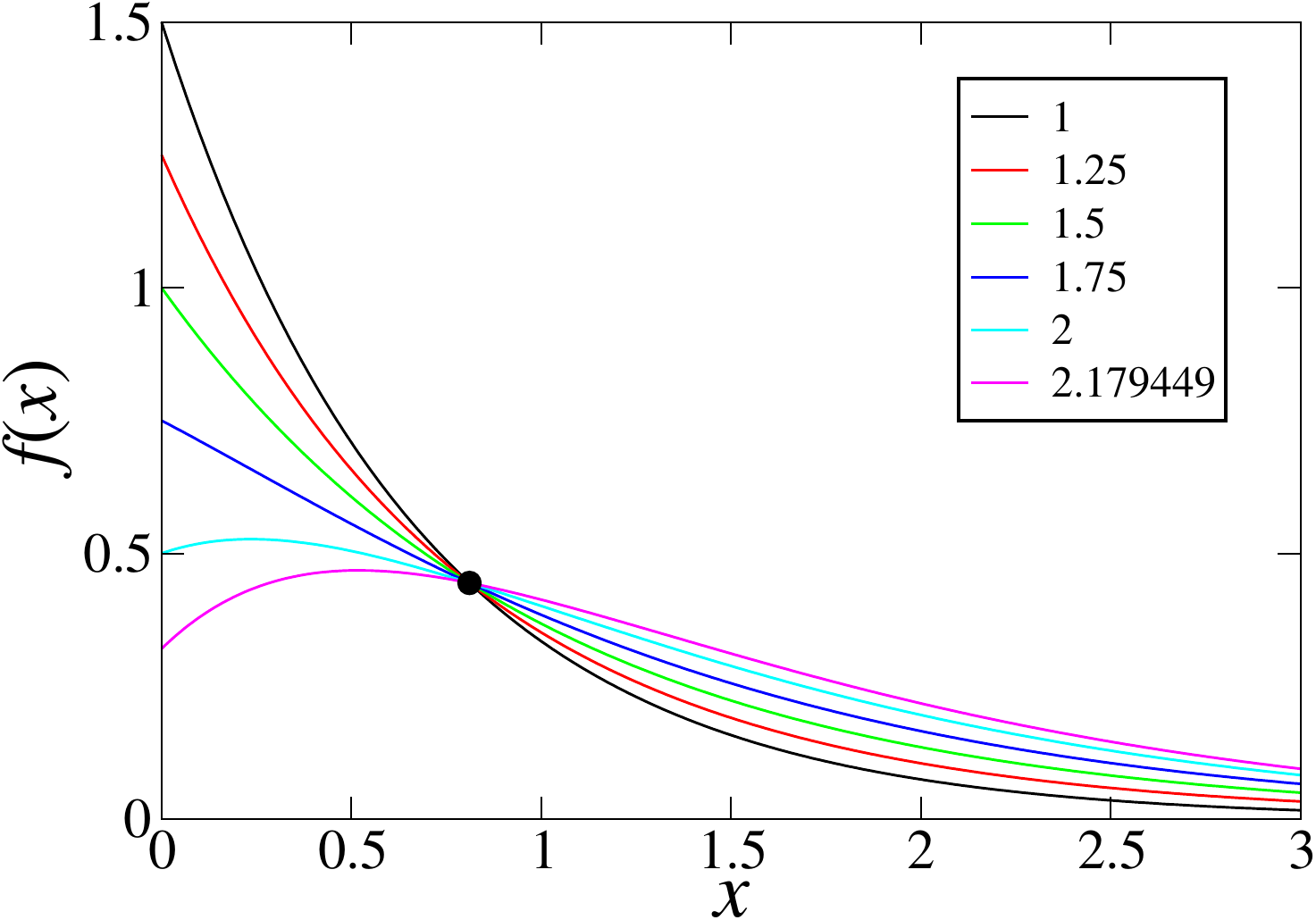}
\caption{
Distribution $f(x)$ of the first positive position, as given by~(\ref{ftwo}),
for the same parameter values as figure~\ref{rhotwoplot},
yielding $f(x_\star)=4/9$.}
\label{ftwoplot}
\end{center}
\end{figure}

Figures~\ref{rhotwoplot} and~\ref{ftwoplot} show the distributions $\rho(x)$ and $f(x)$
for $p_1=1$, $p_2=3/2$ and several $z$ (see legend),
including $z_\max=\sqrt{19}/2\approx2.179449$, for which $\rho(0)=0$.
The expressions~(\ref{rhotwo}) and~(\ref{ftwo}) become independent of $z$ for
$p_1\e^{-p_1x}=p_2\e^{-p_2x}$,
i.e.,
\beq
x_\star=\frac{\ln(p_2/p_1)}{p_2-p_1},\qquad
f(x_\star)=2 \rho(x_\star)=\left(\frac{p_1^{p_2}}{p_2^{p_1}}\right)^{1/(p_2-p_1)}.
\label{xfstar}
\eeq

We have
\beq
\ell=\frac{1}{p_1}+\frac{1}{p_2}-\frac{1}{z},\qquad
\omega=p_1+p_2-z,
\label{ellome}
\eeq
hence (see~(\ref{abdef}))
\beq
\Az=1+\frac{(z-p_1)(p_2-z)}{z^2},\qquad
\Bz=1+\frac{(z-p_1)(p_2-z)}{p_1p_2}.
\eeq
Both quantities equal unity for either $z=p_1$ or $z=p_2$,
where~(\ref{rhotwo}) and~(\ref{ftwo}) reduce to single exponentials.
They are larger than unity in the red region ($p_1<z<p_2$)
and smaller than unity in the blue region ($p_2<z<z_\max$).

The dimensionless quantity $\Az$ reaches its minimum,
\beq
\Az=\frac{2\sqrt{3}-1}{3}\approx0.821367,
\label{a2}
\eeq
in the limit $p_2\to p_1$ and for $z=z_\max=p_1\sqrt{3}$.
In this situation, setting $p_1=1$,
the step distribution becomes the symmetric linear-times-exponential distribution
\beq
\rho(x)=\frac{\abs{x}\e^{-\abs{x}}}{2},
\label{rho2}
\eeq
for which~(\ref{ftwo}) becomes
\beq
f(x)=\left(2-\sqrt{3}+(\sqrt{3}-1)x\right)\e^{-x},
\label{ftwolin}
\eeq
in agreement with~\cite{gmsprl,revue,msbook}.

To close, we mention that the formula~(\ref{g1prod}) leads to
\beq
\hat g(p)=\frac{(p+p_1)(p+p_2)}{p(p+z)},
\eeq
so that
\beq
g_\reg(x)=\frac{p_1p_2}{z}+\frac{(z-p_1)(p_2-z)}{z}\,\e^{-z x},
\label{gtwo}
\eeq
whereas the distribution of the equilibrium excess length $E$ is given by (see~(\ref{felap}))
\beq
\hat f_E(p)=\frac{p_1p_2}{z}\,\frac{p+z}{(p+p_1)(p+p_2)},
\eeq
hence
\beq
f_E(x)=\frac{p_1p_2}{z}\,\frac{(z-p_1)\e^{-p_1x}-(z-p_2)\e^{-p_2x}}{p_2-p_1}.
\label{fetwo}
\eeq
In the above formulas, the prefactor $p_1p_2/z$ equals $1/\sqrt{D}$, as should be.
One can also verify from~(\ref{fetwo}) that $\mean{E}=\ell$, consistent with~(\ref{ellome}).

\subsection{Symmetric Erlang distributions}
\label{exfamily}

Consider the family of symmetric Erlang distributions
\beq
\rho(x)=\frac{\abs{x}^{M-1}\e^{-\abs{x}}}{2(M-1)!},
\label{rhon}
\eeq
parametrised by the integer $M\ge1$\footnote{For usual (one-sided) Erlang distributions, the
parameter $M$ is called the shape parameter.}.
The symmetric exponential distribution~(\ref{rho1}) is recovered for $M=1$,
and the linear-times-exponential distribution~(\ref{rho2}) for $M=2$.
The $M\to\infty$ limit is singular,
as the limiting step density consists of two symmetric delta peaks at $\pm1$ in the reduced variable $x/M$.

\begin{figure}[!htbp]
\begin{center}
\includegraphics[angle=0,width=.6\linewidth,clip=true]{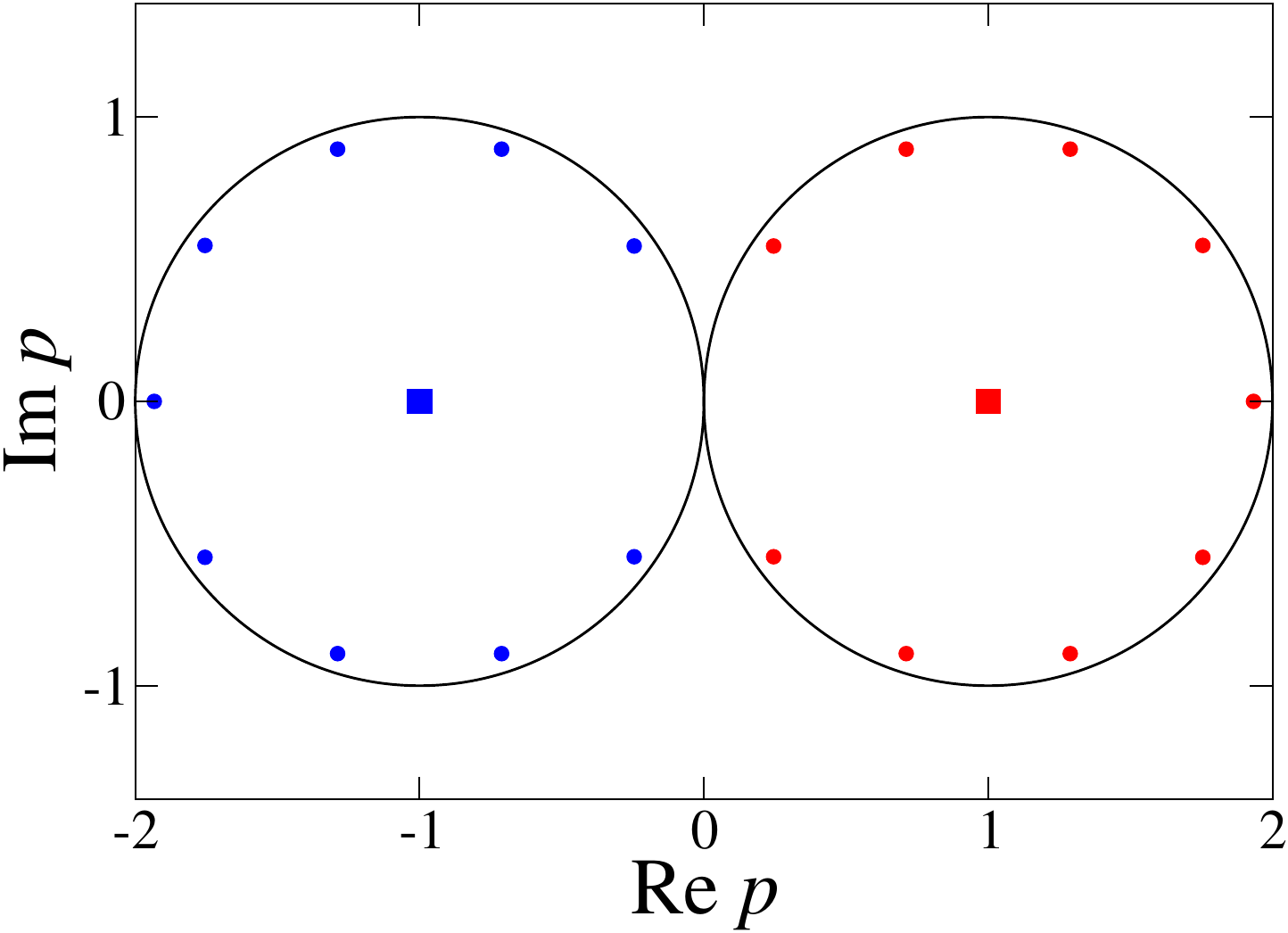}
\caption{
The 9 zeros $z_b$ (red) and their opposites (blue) in the complex $p$-plane for $M=10$.
The circles have unit radii and are centered at $p=\pm1$,
with $p=1$ (red square) being the common value of the decay rates,
and $p=-1$ (blue square) its opposite.}
\label{zerosplot}
\end{center}
\end{figure}

The distribution~(\ref{rhon}) represents a peculiar, maximally degenerate form of~(\ref{rhorat}), obtained by letting the $M$ decay rates $p_a$ simultaneously approach unity, while the corresponding weights $w_a$ diverge appropriately, thereby generalising the construction of the distribution~(\ref{rho2}) for $M=2$.
We have
\beq
\phi(p)=1-\frac12\left(\frac{1}{(1+p)^M}+\frac{1}{(1-p)^M}\right),
\label{phin}
\eeq
and so
\beq
D=\frac{M(M+1)}{2}.
\label{dn}
\eeq
Expanding~(\ref{phin}), we find that the $M-1$ zeros $z_b$ of $\phi(p)$ and their opposites satisfy
a polynomial equation of degree $2M-2$, namely
\beq
\sum_{k=1}^M\left((-1)^k{M\choose k}-{M\choose 2k}\right)p^{2k-2}=0.
\eeq
Figure~\ref{zerosplot} shows a plot of these zeros in the complex $p$-plane for $M=10$.
Even for the modest value $M=10$,
the zeros sit near the circles with unit radii centered at $p=\pm1$.
More details will be given below.
These zeros have appeared in several earlier works involving the family of symmetric Erlang
distributions~(\ref{rhon}).
They are studied in detail in~\cite{lfn},
which focuses on the thermodynamics of random-field Ising chains.
Additionally, a recent work~\cite{BMS} explores the statistics of gaps in random walks with the
same family of step distributions.

The distribution of the first positive position $H$ reads, using~(\ref{f1prod}),
\beq
f(x)=\int\frac{\dd p}{2\pi\ii}\,\e^{p x}\left(1-\frac{p}{(p+1)^M}\,{\prod_b}'(p+z_b)\right).
\eeq
Setting $p=\eps-1$, this becomes
\beq
f(x)=\e^{-x}\int\frac{\dd\eps}{2\pi\ii}\,\e^{\eps
x}\left(1+\frac{1-\eps}{\eps^M}\,{\prod_b}'(\eps+z_b-1)\right).
\eeq
The integrand has a multiple pole of order $M$ at $\eps=0$.
Evaluating the corresponding residue yields
\beq
f(x)=\e^{-x}\sum_{k=0}^{M-1}(S_k-S_{k+1})\frac{x^k}{k!}.
\label{fn}
\eeq
This expression for $f(x)$ has the same structure as the step density~(\ref{rhon}).
It is the product of the decaying exponential $\e^{-x}$ by a polynomial in $x$ of degree $M-1$.
The coefficients of the latter polynomial involve the elementary symmetric polynomials $S_k$ of the
variables $z_b-1$.
These are defined by expanding the product
\beq
{\prod_b}'(\eps+z_b-1)=\sum_{k=0}^{M-1}S_k\,\eps^{M-1-k}.
\eeq
We have
\beq
S_0=1,\qquad
S_1={\sum_b}'(z_b-1),\quad
\dots,\quad
S_{M-1}={\prod_b}'(z_b-1),\quad
S_M=0.
\eeq
Setting $x=0$ in~(\ref{fn}), we obtain
\beq
f(0)=S_0-S_1=M-{\sum_b}'z_b,
\eeq
in agreement with the expression~(\ref{omegasum}) of $\omega$.
The asymptotic decay of $f(x)$
follows the proportionality predicted in~(\ref{fas2}),
with
\beq
K=2S_{M-1}=2\,{\prod_b}'(z_b-1).
\eeq

The distribution $f_E(x)$ of the recurrence length $E$ assumes a form
similar to the expression~(\ref{fn}) of $f(x)$.
The formula~(\ref{feint}) yields
\beq
f_E(x)=\frac{\e^{-x}}{\sqrt{D}}\sum_{k=0}^{M-1}S_k\frac{x^k}{k!},
\label{fen}
\eeq
where $D$ is given by~(\ref{dn}).

For the first three values of the integer $M$, the distribution of $H$ can be made explicit:

\begin{enumerate}

\item[1.]
For $M=1$,
the symmetric exponential distribution~(\ref{rho1}) is recovered.
The set of zeros is empty,
so that~(\ref{fn}) reduces to~(\ref{f1}).

\item[2.]
For $M=2$,
the symmetric linear-times-exponential distribution~(\ref{rho2}) is recovered.
There is a real zero at $z_1=\sqrt{3}$, and so $S_1=\sqrt{3}-1$,
so that~(\ref{fn}) becomes~(\ref{ftwolin}).

\item[3.]
For $M=3$,
there is a pair of conjugate complex zeros, $z_1=a+\ii b$ and $z_2=a-\ii b$, with
\beq
a=\frac{\sqrt{2\sqrt6+3}}{2}\approx1.405256,\qquad
b=\frac{\sqrt{2\sqrt6-3}}{2}\approx0.689017.
\eeq
The expression~(\ref{fn}) then reads
\beq
f(x)=\Bigl(3-2a+(4a-\sqrt6-3)x+(\sqrt6+1-2a)\frac{x^2}{2}\Bigr)\e^{-x}.
\eeq

\end{enumerate}

\begin{figure}[!htbp]
\begin{center}
\includegraphics[angle=0,width=.6\linewidth,clip=true]{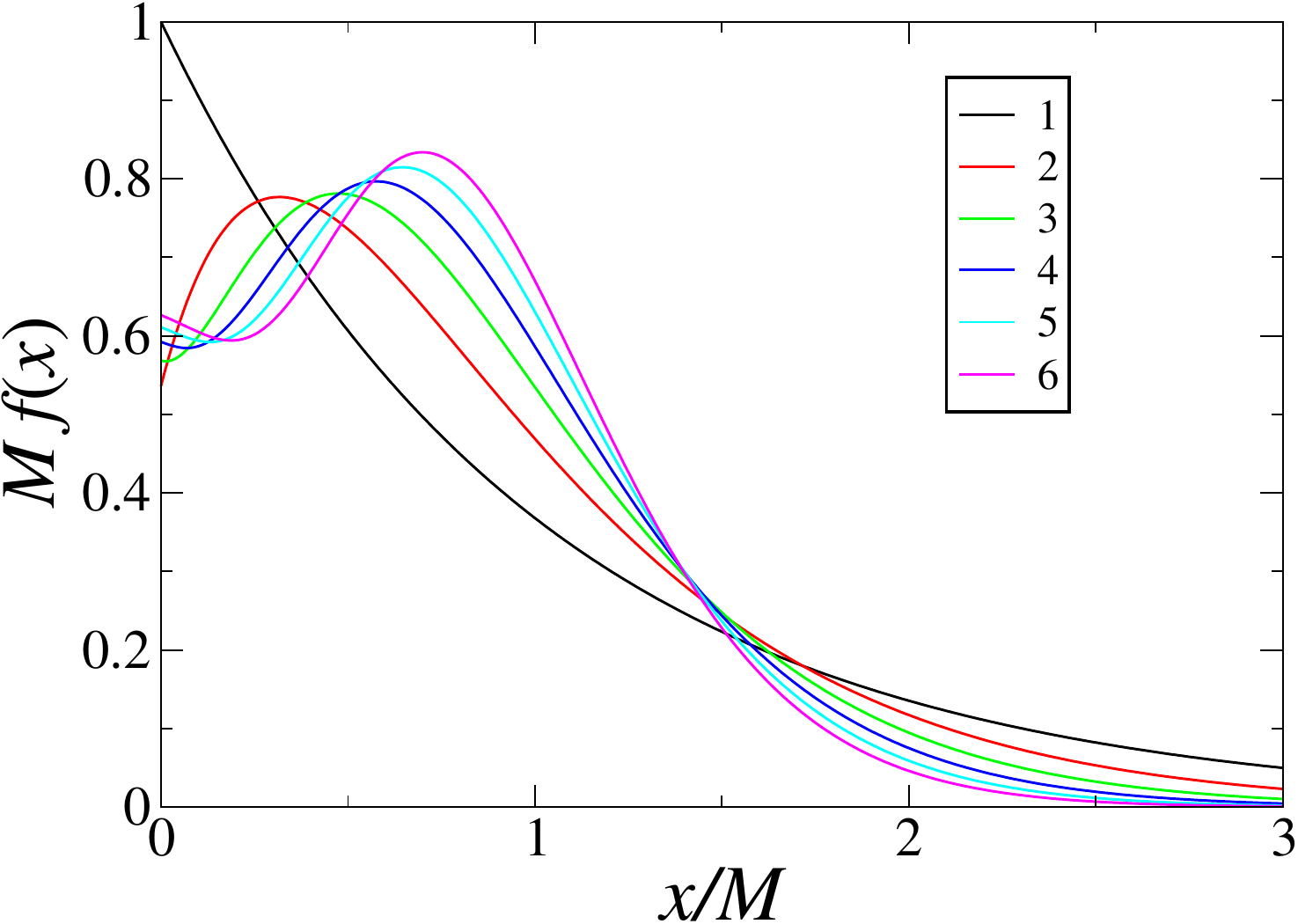}
\caption{
Distribution of $H$ for the first six values of the integer $M$.
The product $M f(x)$ is plotted against the ratio $x/M$.}
\label{nplot}
\end{center}
\end{figure}

Figure~\ref{nplot} shows the distribution of $H$ for the first six values of $M$.
The product $M f(x)$ is plotted against the ratio $x/M$.
This rescaling ensures that the plotted curves have a mild dependence on $M$.

\begin{figure}[!htbp]
\begin{center}
\includegraphics[angle=0,width=.6\linewidth,clip=true]{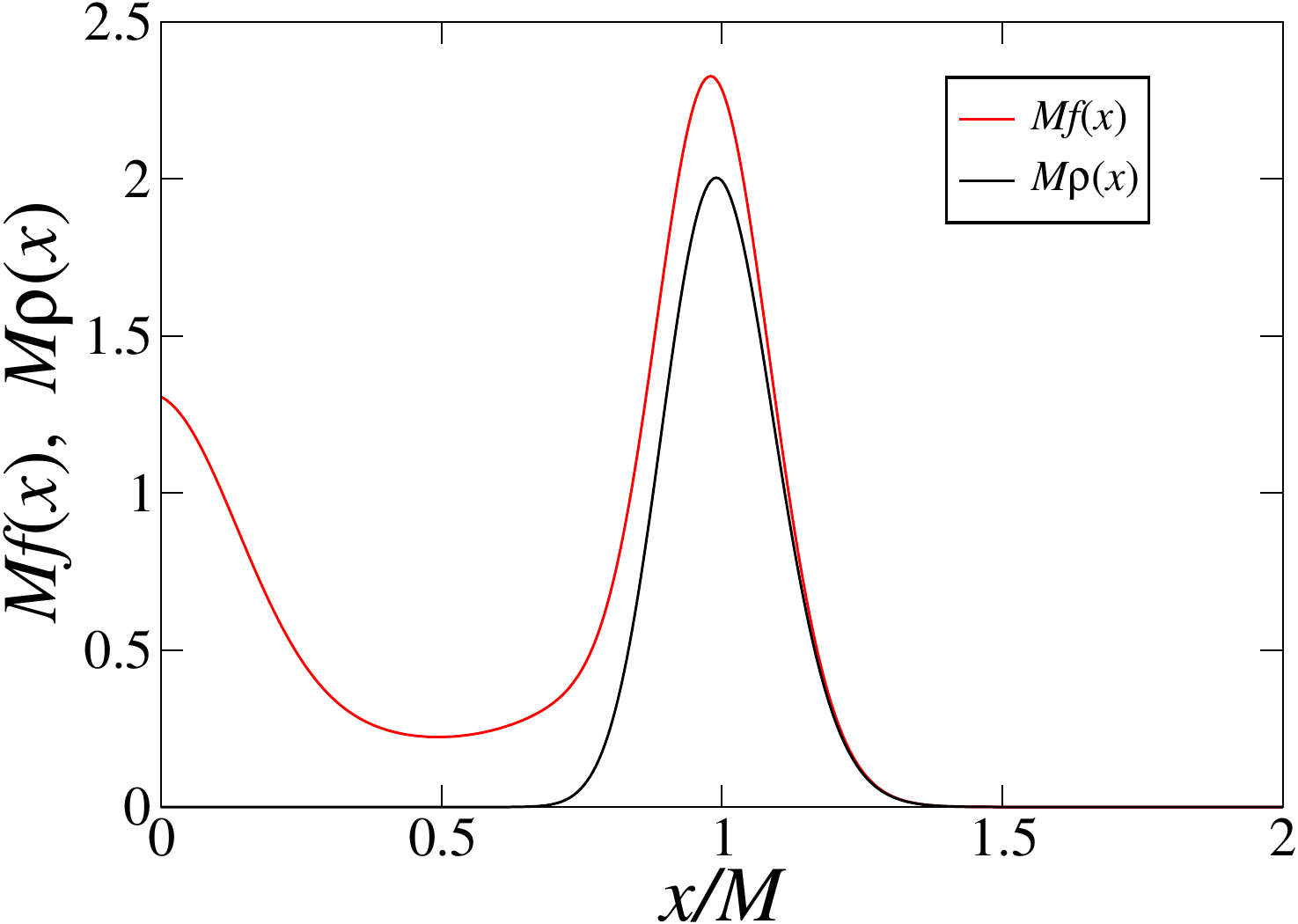}
\caption{
Comparison between the distribution $f(x)$ of the first positive position (red curve)
and the positive part of the step distribution $\rho(x)$ (black curve) for $M=100$.
The products $M f(x)$ and $M\rho(x)$ are plotted against the ratio $x/M$.}
\label{n100plot}
\end{center}
\end{figure}

The remainder of this section is devoted to an analysis
of the crossover to the symmetric binary distribution which takes place at large $M$.
Figure~\ref{n100plot} shows a comparison between the step distribution $\rho(x)$ and the
distribution $f(x)$ of the first positive position for $M=100$.
The products $M \rho(x)$ and $M f(x)$ are again plotted against the ratio $x/M$.
As expected, $\rho(x)$ exhibits a peak around $x/M=1$
(and a symmetric one around $x/M=-1$, but only positive values of $x$ are shown).
More surprisingly, the distribution $f(x)$ exhibits a bimodal structure,
with a first peak in the region where $x\ll M$
and a second one around $x/M=1$, closely resembling that of $\rho(x)$.
Both peaks are separated by a pronounced dip.

The above features are corroborated by the following analysis
of the first few moments of $H$ (see~(\ref{moms})) at large $M$.
The diffusion coefficient is given by~(\ref{dn}),
whereas~(\ref{ceven}) yields
\beq
c_2=\frac{(M+2)(M+3)}{12}.
\label{ell2n}
\eeq
The extrapolation length (see~(\ref{ellsum}))
\beq
\ell=M-{\sum_b}'\frac{1}{z_b}
\label{elln}
\eeq
deserves some more care.
The asymptotic behaviour of this quantity,
and of related sums and products over the zeros $z_b$,
has been thoroughly studied in~\cite{lfn}.
To leading order, as $M$ becomes large,
the zeros are uniformly distributed on the circle centered at $p=1$ with a radius close to unity
(see~figure~\ref{zerosplot}), and more precisely given by $2^{-1/M}$.
We choose to number the zeros so as to have
\beq
z_b\approx1-2^{-1/M}\e^{-2\pi\ii b/M}\qquad(b=1,\dots,M-1).
\label{zuni}
\eeq
For large but finite $M$,
the correction to the uniform circular distribution~(\ref{zuni}) is most significant for the zeros
located near the origin,
i.e., either $b\ll M$ or $M-b\ll M$.
The leading correction assumes the following scaling form~\cite[Sec.~6.2]{lfn}.
Setting
\beq
\xi=\frac{2\pi b}{\sqrt{M}},
\eeq
we have
\beq
z_b\approx\frac{2\pi\ii b+Y(\xi)}{M},
\qquad Y(\xi)=\frac{\xi^2}{2}+\ln\left(1+\sqrt{1-\e^{-\xi^2}}\right).
\label{zsca}
\eeq
The scaling function $Y(\xi)$ is real, and therefore only affects the real parts of the zeros.
Inserting~(\ref{zsca}) into~(\ref{elln}), we obtain after some algebra
\beq
\ell\approx\frac{M}{2}\left(1-\frac{1}{\pi\sqrt{M}}\ln\frac{M}{M_0}\right),
\label{ellsca}
\eeq
with
\beq
M_0
=4\pi^2\exp\left(-2\int_0^\infty\frac{\dd
\xi}{\xi^2}\,\left(Y(\xi)-\frac{\xi^2}{2}-\xi\e^{-\xi}\right)\right)
\approx8.166752.
\eeq
The difference $M-\ell$ is denoted by $\Omega_{(-1)}$ in~\cite{lfn},
and the expression~(\ref{ellsca}) corresponds to equation~(6.37) therein.

Inserting the expressions~(\ref{dn}),~(\ref{ellsca}) and~(\ref{ell2n})
of $D$, $c_1=\ell$ and $c_2$ into~(\ref{moms}),
we obtain the following asymptotic formulas for the first three moments of $H$ at large~$M$,
where relative corrections of order $1/M$ are neglected:
\beqa
\mean{H}&\approx&\frac{M}{\sqrt2},
\nonumber\\
\mean{H^2}&\approx&\frac{M^2}{\sqrt2}\left(1-\frac{1}{\pi\sqrt{M}}\ln\frac{M}{M_0}\right),
\nonumber\\
\mean{H^3}&\approx&\frac{M^3}{\sqrt2}\left(1-\frac{3}{2\pi\sqrt{M}}\ln\frac{M}{M_0}\right).
\label{momsn}
\eeqa
The leading-order estimates imply that the peaks observed in figure~\ref{n100plot} for $M=100$
indeed become asymptotically well separated,
and that the weights $W_1$ and $W_2$ of the first and second peak respectively read
\beq
W_1=1-\frac{1}{\sqrt2}\approx0.292893,\qquad
W_2=\frac{1}{\sqrt2}\approx0.707107.
\eeq
The negative corrections in $(\ln M)/\sqrt{M}$ occurring in the expressions~(\ref{momsn}) of
$\mean{H^2}$ and $\mean{H^3}$ most probably affect higher-order moments as well.

\begin{figure}[!htbp]
\begin{center}
\includegraphics[angle=0,width=.6\linewidth,clip=true]{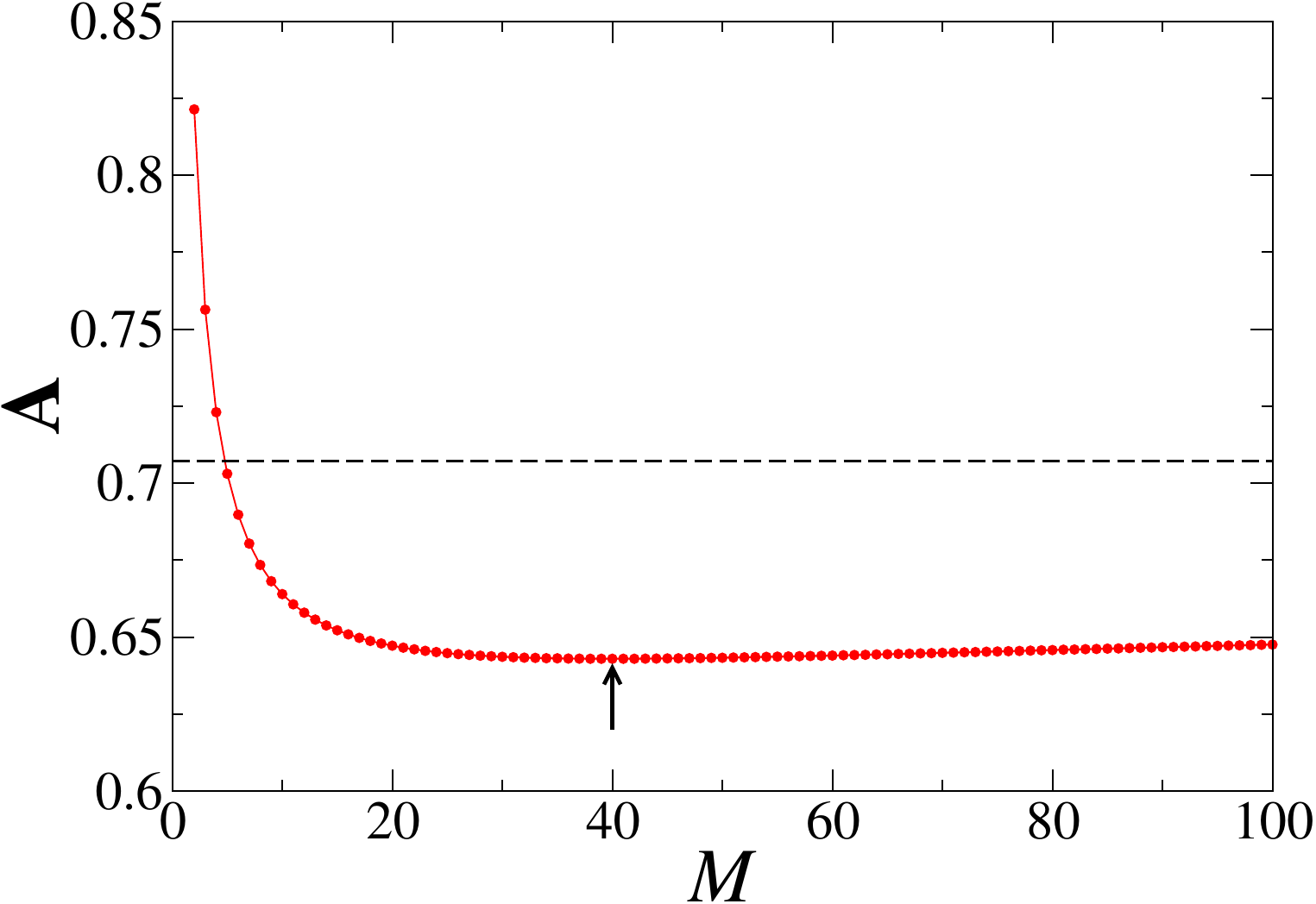}
\caption{
Dimensionless quantity $\Az$ defined in~(\ref{abdef}) plotted against the integer $M$ between 2 and
100.
Dashed horizontal line: limit $\Az_\infty=1/\sqrt2$.
Arrow: minimum $\Az\approx0.642960$ reached for $M=40$.}
\label{aplot}
\end{center}
\end{figure}

In terms of the dimensionless quantity $\Az$ defined in~(\ref{abdef}),
the asymptotic expression~(\ref{ellsca}) translates to
\beq
\Az\approx\frac{1}{\sqrt2}\left(1-\frac{1}{\pi\sqrt{M}}\ln\frac{M}{M_0}\right).
\label{asca}
\eeq
Figure~\ref{aplot} shows that $\Az$ has a non-monotonic dependence on the integer $M$,
dropping fast from $\Az=1$ at $M=1$ (out of scale) to $\Az\approx0.821367$ for $M=2$
(see~(\ref{a2})),
reaching a minimum,
\beq
\Az\approx0.642960,
\label{a40}
\eeq
for $M=40$ (arrow),
and going up very slowly to its limit $\Az_\infty=1/\sqrt2$ (dashed line),
according to~(\ref{asca}).

The quantity $\Az$ has been shown to provide a measure of the dispersion of the distribution of $H$
(see~(\ref{varh}))
and to satisfy the inequality $\Az>1/2$.
It seems quite plausible that $\Az$ reaches a non-trivial absolute minimum $\Az_\min$
for some well-defined step distribution.
An exploratory numerical study supports this hypothesis
and suggests that $\Az_\min$ is only slightly below the value given in~(\ref{a40}).
The exact determination of $\Az_\min$ and of the parent step distribution is left as a challenging
open problem.

\subsubsection*{Author contribution statement}

\small
Both authors contributed equally to the present work,
were equally involved in the preparation of the manuscript,
and have read and approved the final manuscript.

\subsubsection*{Data availability statement}

\small
Data sharing not applicable to this article.

\subsubsection*{Conflict of interest}

\small
The authors declare no conflict of interest.

\begin{appendix}

\section{Details of derivations}

\subsection{Derivation of equation~(\ref{eq:smith})}
\label{appsmith}

Reasoning along the lines of~\cite{glrenew}, it is easy to obtain the probability density of
$\Sigma_{R_x+1}$ in Laplace space.
The double Laplace transform of this density with respect to its two arguments (space $x$ and value
$y$ of the random variable) reads
\be
\lap{x,y}f_{\Sigma_{R_x+1}}(x,y)
=\lap{x}\mean{\e^{-u\Sigma_{R_x+1}}}
=\frac{\hat f(u)-\hat f(p+u)}{p(1-\hat f(p+u))}.
\ee
Taking the derivative of minus this expression with respect to $u$ at $u=0$ gives
\be
\lap{x}\mean{\Sigma_{R_x+1}}
=\frac{\mean{H}}{p(1-\hat f(p))}.
\ee
The right-hand side of this equation equals the Laplace transform with respect to $x$ of
the right-hand side of~(\ref{eq:smith})
because (see, e.g.,~\cite[Eq.~(3.4)]{glrenew})
\beq\label{eq:meanN}
\lap{x}\mean{R_x}
=\frac{\hat f(p)}{p(1-\hat f(p))}.
\eeq
Note that $\mean {\Sigma_{R_x}}$ is not equal to $\mean{H}\mean{R_x}$.

\subsection{Proof that~(\ref{eq:lemma1}) implies the Pollaczek-Spitzer formula}
\label{appequiv}

Recall~(\ref{eq:lemma1})
\be
\ln\frac{1}{1-\til f(s,q)}=\sum_{n\ge1}\frac{s^n}{n}\int_{0}^\infty\dd x\,\e^{\ii q
x}f_{x_n}(x)=J(q).
\ee
Consider
\be
I(p)=\int_{0}^\infty\dd x\,\e^{-px}\sum_{n\ge1}\frac{s^n}{n}f_{x_n}(x)
\qquad {\Re (p)>0},
\ee
which is such that $J(q)=I(-\ii q)$.
The goal is to show that $I(p)$ is given by~(\ref{eq:resIp}), which is the integral appearing in
the Pollaczek-Spitzer formula~(\ref{eq:Poll-S1}) or~(\ref{eq:ps}) (see also~\eqref{ione}).
The proof goes as follows.
Rewrite
\beqa
\sum_{n\ge1}\frac{s^n}{n}f_{x_n}(x)&=&\sum_{n\ge1}\frac{s^n}{n}\int_{-\infty}^\infty\frac{\dd
q}{2\pi}\e^{\ii qx}\til \rho(q)^n
\nonumber\\
&=&-\int_{-\infty}^\infty\frac{\dd q}{2\pi}\e^{\ii qx}\ln(1-s\til \rho(q)).
\eeqa
Thus
\beqa
I(p)&=&-\int_{0}^\infty\dd x\,\e^{-px}\int_{-\infty}^\infty\frac{\dd q}{2\pi}\e^{\ii qx}\ln(1-s\til
\rho(q))
\nonumber\\
&=&-\int_{-\infty}^\infty\frac{\dd q}{2\pi}\ln(1-s\til \rho(q))\int_{0}^\infty\dd
x\,\e^{-px}\e^{\ii qx}
\nonumber\\
&=&-\int_{-\infty}^{\infty}\frac{\dd q}{2\pi}\frac{\ln(1-s\til \rho(q))}{p-\ii q}.
\eeqa
The integral is split into two parts
\be
-I(p)=\int_{-\infty}^{0}\frac{\dd q}{2\pi}\frac{\ln(1-s\til \rho(q))}{p-\ii q}
+\int_{0}^{\infty}\frac{\dd q}{2\pi}\frac{\ln(1-s\til \rho(q))}{p-\ii q},
\ee
yielding finally
\beq\label{eq:resIp}
I(p)=-\frac{p}{\pi}\int_{0}^{\infty}\dd q\,\frac{\ln(1-s\til \rho(q))}{p^2+q^2}.
\eeq

\subsection{Derivation of equation~(\ref{codd})}
\label{appsub}

We start by simplifying notations by setting (see~\eqref{eq:coeff})
\beq
A(q)=\ln\frac{1-\til\rho(q)}{Dq^2}=\sum_{n\ge1}a_{2n}q^{2n},\qquad
a_{2n}=2(-1)^n\,\frac{c_{2n}}{(2n)!}.
\label{appnots}
\eeq
The Mellin transform
\beq
\mu_K(s)=\int_0^\infty\dd q\,q^{s-1}\,A(q)
\label{muapp}
\eeq
is convergent for $-2<\Re s<0$.
It has a meromorphic continuation in the whole left-hand half-plane ($\Re s<0$),
with poles at $s=-2m$ for $m=1,2,\dots$
We are interested in the values $\mu_K(-2m-1)$ for $m=0,1,\dots$,
which enter~(\ref{muodd}).
These quantities can be derived by splitting the definition~(\ref{muapp}) as
\beq
\mu_K(s)=\mu_{K,1}(s)+\mu_{K,2}(s),
\eeq
with
\beqa
\mu_{K,1}(s)&=&\int_0^1\dd q\,q^{s-1}\,A(q)\qquad(\Re s>-2),
\label{mu1}
\\
\mu_{K,2}(s)&=&\int_1^\infty\dd q\,q^{s-1}\,A(q)\qquad(\Re s<0).
\label{mu2}
\eeqa

We henceforth fix the value of the integer $m$.
The expression~(\ref{mu1}) can be recast as
\beqa
\mu_{K,1}(s)
&=&\int_0^1\dd q\,q^{s-1}\biggl(A(q)-\sum_{n=1}^ma_{2n}q^{2n}\biggr)
+\int_0^1\dd q\,q^{s-1}\sum_{n=1}^ma_{2n}q^{2n}
\nonumber\\
&=&\int_0^1\dd q\,q^{s-1}\biggl(A(q)-\sum_{n=1}^ma_{2n}q^{2n}\biggr)
+\sum_{n=1}^m\frac{a_{2n}}{s+2n}.
\eeqa
The integral entering the above expressions is convergent for $-2m-2<\Re s<-2m$.
We have in particular
\beq
\mu_{K,1}(-2m-1)
=\int_0^1\frac{\dd q}{q^{2m+2}}\biggl(A(q)-\sum_{n=1}^ma_{2n}q^{2n}\biggr)
-\sum_{n=1}^m\frac{a_{2n}}{2m-2n+1}.
\label{mu1res}
\eeq
The corresponding expression for $\mu_{K,2}(s)$ can be directly evaluated at $s=-2m-1$:
\beqa
\mu_{K,2}(-2m-1)
&=&\int_1^\infty\frac{\dd q}{q^{2m+2}}\biggl(A(q)-\sum_{n=1}^ma_{2n}q^{2n}\biggr)
+\int_1^\infty\frac{\dd q}{q^{2m+2}}\sum_{n=1}^ma_{2n}q^{2n}
\nonumber\\
&=&\int_1^\infty\frac{\dd q}{q^{2m+2}}\biggl(A(q)-\sum_{n=1}^ma_{2n}q^{2n}\biggr)
+\sum_{n=1}^m\frac{a_{2n}}{2m-2n+1}.
\label{mu2res}
\eeqa
Summing up~(\ref{mu1res}) and~(\ref{mu2res}) yields
\beq
\mu_K(-2m-1)=\int_0^\infty\frac{\dd q}{q^{2m+2}}\biggl(A(q)-\sum_{n=1}^ma_{2n}q^{2n}\biggr).
\label{melsub}
\eeq
Inserting this expression into~(\ref{muodd}), using the notations~(\ref{appnots}), we
obtain~(\ref{codd}).

\subsection{Derivation of equation~(\ref{gprod})}
\label{appg}

\begin{enumerate}

\item[1.]
In the definition~(\ref{glapdef}) of the Laplace transform of $g(s,x)$,
\beq
\hat g(s,p)=\int_0^\infty\dd x\, g(s,x)\,\e^{-p x}
\eeq
for $\Re p>0$,
replace $g(s,x)$ by the right-hand side of the linear equation~(\ref{gmilne}), obtaining
\beq
\hat g(s,p)=1+s\int_0^\infty\dd x\,\e^{-p x}\int_0^\infty\dd y\, g(s,y)\, \rho(x-y).
\label{app1}
\eeq

\item[2.]
In~(\ref{app1}) express $g(s,y)$ in terms of its Laplace transform,
\beq
g(s,y)=\int\frac{\dd q}{2\pi\ii}\,\hat g(s,q)\,\e^{q y},
\eeq
where the integration contour is vertical with $\Re q>0$,
and similarly for $\rho(x-y)$, obtaining
\beq
\hat g(s,p)=1
+s\int_0^\infty\dd x\,\e^{-p x}\int_0^\infty\dd y
\int\frac{\dd q}{2\pi\ii}\,\hat g(s,q)\,\e^{q y}
\int\frac{\dd r}{2\pi\ii}\,\hat\rho(r)\,\e^{r(x-y)}.
\label{app2}
\eeq

\item[3.]
In~(\ref{app2}) perform the integrations over $x$ and $y$, obtaining
\beq
\hat g(s,p)=1+s
\int\frac{\dd q}{2\pi\ii}\,\hat g(s,q)
\int\frac{\dd r}{2\pi\ii}\,\frac{\hat\rho(r)}{(p-r)(r-q)}
\label{app3}
\eeq
for $0<\Re q<\Re r<\Re p$ and $\Re r<p_1$,
with $p_1$ being the smallest decay rate entering~(\ref{rhorat}).

\item[4.]
In~(\ref{app3}) shift the $q$-contour to the right.
The contribution of the pole at $q=r$ yields
\beq
\hat g(s,p)=1+s
\int\frac{\dd r}{2\pi\ii}\,\frac{\hat g(s,r)\hat\rho(r)}{p-r}
\label{app4}
\eeq
for $0<\Re r<\Re p$ and $\Re r<p_1$.

\item[5.]
In~(\ref{app4}) shift the $r$-contour to the right of the pole at $r=p$, obtaining
\beq
\hat g(s,p)=1+s\hat g(s,p)\hat\rho(p)
+s\int\frac{\dd r}{2\pi\ii}\,\frac{\hat g(s,r)\hat\rho(r)}{p-r},
\eeq
i.e., using the definition~(\ref{phidef}) of $\phi(s,p)$,
\beq
\phi(s,p)\hat g(s,p)=1+s
\int\frac{\dd r}{2\pi\ii}\,\frac{\hat g(s,r)\hat\rho(r)}{p-r},
\label{app5}
\eeq
for $0<\Re p<\Re r<p_1$.

\item[6.]
In~(\ref{app5}) shift the $r$-contour to the right.
The contributions of the poles at $r=p_a$ yield an expression of the form
\beq
\phi(s,p)\hat g(s,p)=1+\sum_a\frac{C_a}{p-p_a},
\label{app6}
\eeq
where the $C_a$ are constants.

\item[7.]
This last stage is the gist of the factorisation technique, which works as follows in the present
setting.
The Laplace transform $\hat g(s,p)$ is analytic for $\Re p>0$.
The zeros of $\phi(s,p)$ at $p=z_b$ must therefore be zeros of the right-hand side of~(\ref{app6}).
Moreover, this right-hand side goes to unity as $p\to+\infty$.
We are thus left with the product formula
\beq
\phi(s,p)\hat g(s,p)=\frac{\prod_b(p-z_b)}{\prod_a(p-p_a)},
\eeq
or equivalently
\beq
\hat g(s,p)=\frac{\prod_a(p+p_a)}{\prod_b(p+z_b)},
\label{app7}
\eeq
which is~(\ref{gprod}).

\end{enumerate}

\subsection{Derivation of equation~(\ref{fprod})}
\label{appf}

\begin{enumerate}
\item[1.]
In the expression of the Laplace transform of $f(s,x)$,
\beq
\hat f(s,p)=\int_0^\infty\dd x\, f(s,x)\,\e^{-p x}
\eeq
for $\Re p>0$,
replace $f(s,x)$ by the right-hand side of its expression~(\ref{fgmilne}), obtaining
\beq
\hat f(s,p)=s\int_0^\infty\dd x\,\e^{-p x} \int_0^\infty\dd y\, g(s,y)\, \rho(x+y).
\label{app8}
\eeq

\item[2.]
In~(\ref{app8}) express $g(s,y)$ and $\rho(x+y)$ in terms of their Laplace transforms, obtaining
\beq
\hat f(s,p)
=s\int_0^\infty\dd x\,\e^{-p x} \int_0^\infty\dd y
\int\frac{\dd q}{2\pi\ii}\,\hat g(s,q)\,\e^{q y}
\int\frac{\dd r}{2\pi\ii}\,\hat\rho(r)\,\e^{-r(x+y)}.
\label{app9}
\eeq

\item[3.]
In~(\ref{app9}) perform the integrations over $x$ and $y$, obtaining
\beq
\hat f(s,p)=s
\int\frac{\dd q}{2\pi\ii}\,\hat g(s,q)
\int\frac{\dd r}{2\pi\ii}\,\frac{\hat\rho(r)}{(p+r)(r-q)}
\label{app10}
\eeq
for $0<\Re q<\Re r<p_1$,
with $p_1$ being the smallest decay rate entering~(\ref{rhorat}).

\item[4.]
In~(\ref{app10}) shift the $q$-contour to the right of the pole at $q=r$, obtaining
\beq
\hat f(s,p)=s\int\frac{\dd r}{2\pi\ii}\,\frac{\hat g(s,r)\hat\rho(r)}{p+r}
\label{app11}
\eeq
for $0<\Re r<p_1$.
Using~(\ref{phidef}) and~(\ref{ggprod}),
the integrand can be recast as
\beqa
s\hat g(s,r)\hat\rho(r)
&=&(1-\phi(s,r))\hat g(s,r)
\nonumber\\
&=&\hat g(s,r)-\frac{1}{\hat g(s,-r)}
=\left(\hat g(s,r)-1\right)+\left(1-\frac{1}{\hat g(s,-r)}\right).
\label{app12}
\eeqa

\item[5.]
In~(\ref{app11}) replace the integrand by the last line of~(\ref{app12}), obtaining
\beq
\hat f(s,p)
=\int\frac{\dd r}{2\pi\ii}\,\frac{\hat g(s,r)-1}{p+r}
+\int\frac{\dd r}{2\pi\ii}\left(1-\frac{1}{\hat g(s,-r)}\right)\frac{1}{p+r}.
\label{app13}
\eeq
In the first integral, shift the $r$-contour to the right:
the integral vanishes.
In the second integral, shift the $r$-contour to the left.
The contribution of the pole at $r=-p$ yields
\beq
\hat f(s,p)=1-\frac{1}{\hat g(s,p)},
\eeq
which is~(\ref{fprod}).

\end{enumerate}

\end{appendix}

\bibliography{paperWH.bib}

\end{document}